\documentclass[%
reprint,
superscriptaddress,
showkeys,
twoside,
amsmath,
amssymb,
aps,
prd,
]{revtex4-2}
\usepackage{mathtools}
\usepackage{xspace}
\usepackage{siunitx}
\usepackage{journals}
\usepackage{multirow}
\usepackage{xcolor}
\usepackage[pdftex, pdftitle={Article}, pdfauthor={Author}, hidelinks]{hyperref}
\usepackage{tensor}
\usepackage{cancel}
\usepackage{subfig}
\usepackage[]{graphicx}%draft
\usepackage[utf8]{inputenc}

\DeclareMathOperator{\cov}{cov}
\DeclareMathOperator{\dCor}{dCor}
\DeclareMathOperator{\dVar}{dVar}
\DeclareMathOperator{\dCov}{dCov}
\DeclareMathOperator{\MIC}{MIC}

\DeclareMathOperator{\TP}{TP}
\DeclareMathOperator{\FN}{FN}
\DeclareMathOperator{\FP}{FP}
\DeclareMathOperator{\TN}{TN}
\DeclareMathOperator{\TPR}{TPR}
\DeclareMathOperator{\PPV}{PPV}
\DeclareMathOperator{\FPR}{FPR}

\newcommand{\AU}{universal\xspace}
\usepackage{tabularx}

\begin{document}

\begin{abstract}
We present applications of statistical data analysis methods from both bi- and multivariate statistics to find suitable sets of neutron star features that can be leveraged for accurate and EoS independent -- or \emph{universal} -- relations. To this end, we investigate the ability of various correlation measures such as Distance Correlation and Mutual Information in identifying universally related pairs of neutron star features. We also evaluate relations produced by methods of multivariate statistics such as Principal Component Analysis to assess their suitability for producing universal relations with multiple independent variables.

As part of our analyses, we also put forward multiple entirely novel relations, including a multivariate relation for the $f$-mode frequency of neutron stars with a reduced average relative error of $\num{0.01}$, compared to an error of $\num{0.015}$ of existing, bivariate relations.
\end{abstract}
%\tableofcontents
\title{Finding Universal Relations using Statistical Data Analysis}
% \AU thor{Praveen Manoharan and Kostas D. Kokkotas}
 \author{Praveen Manoharan}
	\email{praveen.manoharan@uni-tuebingen.de}
	\affiliation{Theoretical Astrophysics, IAAT, University of T\"ubingen, 72076 T\"ubingen, Germany}
 \author{Kostas D. Kokkotas}
    \email{kostas.kokkotas@uni-tuebingen.de}
    \affiliation{Theoretical Astrophysics, IAAT, University of T\"ubingen, 72076 T\"ubingen, Germany}

\date{\today}

\maketitle

\section{Introduction}\label{sec:introduction}
The successful detection of gravitational waves from binary neutron star (BNS) mergers through the LIGO-Virgo detectors~\cite{2017PhRvL.119p1101A,2020ApJ...892L...3A} has opened a new avenue into probing and understanding the structure of neutron stars and will hopefully allow us to eventually uncover their true equation of state (EoS). 

Important tools for this task are EoS independent -- or (approximately) universal -- relations that allow for the inference of neutron star bulk parameters through information extracted from gravitational waves. Inspired by early work on such universal relations for isolated neutron stars~\cite{PhysRevLett.77.4134,Andersson98,PhysRevLett.95.151101,lau2010inferring,chan2014multipolar,Yagi365}, the last five years have also given rise to universal relations for binary neutron stars (BNS)~\cite{PhysRevLett.115.091101,PhysRevD.93.124051,PhysRevD.101.084006}: they relate features of the pre-merger neutron stars to the early post-merger remnant, primarily relying on numerical relativity simulations.

Following our own recent work on universal relations for BNS using perturbative calculations~\cite{2021PhRvD.104b3005M,2021FrASS...8..166K}, we found that, with the increasing number of features and amount of data that theoretical computations are able to produce, the traditional method of relying on physical intuition to find \AU relations might not always uncover all possible or the best \AU relations for a given scenario: instead, an automated approach fueled by statistical data analysis might yield better results in finding highly correlated features, and the best functional form to relate them with. A recent work by Soldateschi et al.~\cite{2021A&A...654A.162S} demonstrated the application of principal component analysis (PCA) to the construction of \AU relations with multiple independent variables. 

In this paper, we present applications of statistical data analysis methods from both bi- and multivariate statistics to find suitable sets of neutron star features that can be leveraged for accurate and EoS independent relations. To this end, we first analyze the effectiveness of four different correlation measures -- Pearson Correlation, Distance Correlation~\cite{DistanceCor}, Mutual Information~\cite{MutualInf} and Maximal Information~\cite{MaxInf} -- in identifying pairs of features amenable to universal relations. We find that the conventional wisdom that Pearson Correlation only detects linearly correlated features also applies to the use case of finding bivariate \AU relations for neutron stars. Furthermore, we also find that mutual information based features are more suited for finding non-linear correlation between features, making them more useful for this application.  

In a second step, inspired by~\cite{2021A&A...654A.162S}, we investigate the application of principal component analysis (PCA) in constructing multivariate \AU relations, i.e. relations with multiple independent variables. We find this method suitable for constructing \AU relations that combine several features of a neutron star to predict a target feature. Among others, we find the an entirely novel relation between the average density $\tilde\rho = \sqrt{M/R^3}$, compactness $C = M/R$ and the $f$-mode frequency $\omega_f$ of a neutron star of the form
\begin{equation}
\omega_f = 0.00017 \hat F^2 + \num{0.00647675} \hat F + \num{0.00276012}
\end{equation}
with 
\begin{equation}
\hat F = \num{6.9109313199999995} \frac{\tilde\rho}{0.04} - \num{1.71649574} \frac{C \tilde\rho}{0.01} \ .
\end{equation}
Since the factor $\hat F$ is approximately proportional to the factor $\left(1-C\right)$, this relation could be considered a first order, relativistic correction to the original relation between $\tilde\rho$ and $\omega_f$ derived by Andersson and Kokkotas~\cite{PhysRevLett.77.4134,Andersson98}, which was inspired from Newtonian gravity. In particular, it can be considered a step towards the well known general relativistic \AU relation between the $f$-mode frequency $\omega_f$ and the compactness $C$ put forward by Tsui and Leung~\cite{PhysRevLett.95.151101}.

We perform our analyses using two different data sets from the literature~\cite{2021PhRvD.104b3005M,Kuan:2021jmk}, exemplifying the generalizability of the methods discussed in this work. The results in this work present a first step towards a automated, statistical data analysis driven effort towards the identification and construction of \AU relations for neutron stars (and other objects of astrophysical interest). In a time where the amounts of theoretical model data for astrophysical objects is drastically increasing, we expect having such robust and automated methods available as tools will have a tremendous effect on the quality and quantity of \AU relations that will become available in the future. 

\medskip
\noindent
\textbf{Outline.} We begin by introducing the two data sets that we will base our analyses on in Section~\ref{sec:data}. We then introduce the bivariate approach to finding \AU relations in Section~\ref{sec:bivariate}, and discuss the found relations and the implications for the effectiveness of the analyzed correlation measures in Section~\ref{sec:bivariate_relations}. 

In Section~\ref{sec:multivariate}, we introduce the multivariate approach based on PCA for finding \AU relations, before we discuss some exemplary \AU relations we were able to construct in Section~\ref{sec:multivariate_relations}. We finally conclude our work and give an outlook into potential future directions in Section~\ref{sec:conclusion}.

Note that, unless stated otherwise, we will assume geometrized units in which $G = c = 1$ throughout this paper.

\begin{figure}[t]
\centering
\includegraphics[width=0.5\textwidth]{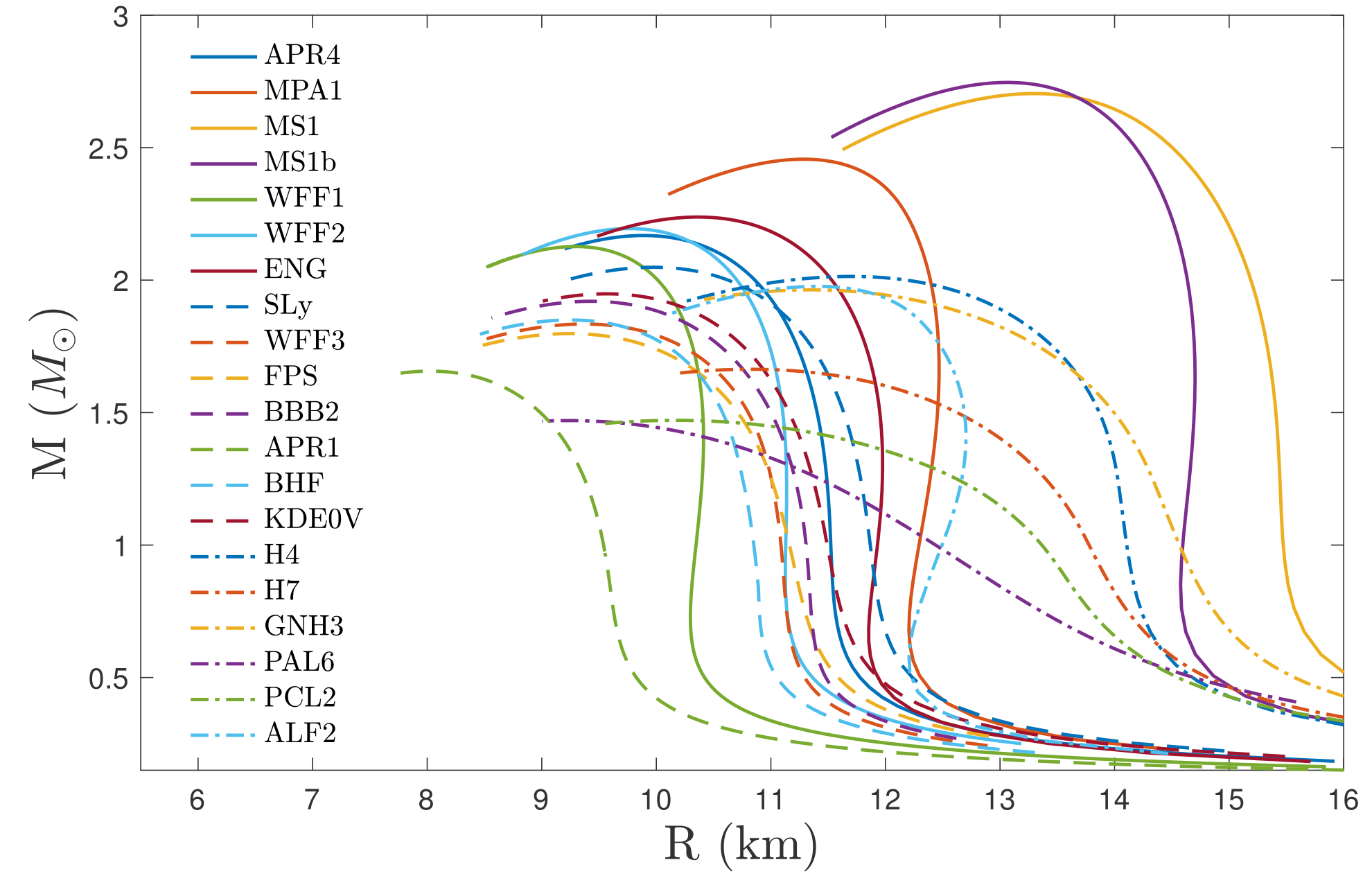}
\caption{Mass-radius relations taken from~\cite[Figure 1]{Kuan:2021jmk}.}
\label{fig:EoS}
\end{figure}
\section{Neutron Star Data}\label{sec:data}
In this work we consider non-rotating neutron stars from a wide range of equations of state. We here give a brief description of the origin and shape of the data sets we utilize for our analysis. For a detailed treatment of the computation of the neutron star models we refer to the original work~\cite{Kuan:2021jmk,2021PhRvD.104b3005M}.
\subsection{Data Sets}\label{sec:data_sets}
For our analyses, we utilize two different data sets that were used in previous publications: \textbf{Data set A} contains a subset of around $58$ non-rotating neutron star models contained in the data set originally put forward in~\cite{kruger2019fast} for the study of rotating and non-rotating neutron stars. This subset covers five different EoSs.  

\textbf{Data set B} contains a subset of $126$ non-rotating neutron star models contained in the data set originally put forward in~\cite{Kuan:2021jmk} for the study of $f$- and $g$-mode frequencies of non-rotating neutron stars. This subset covers a wider range of 15 EoSs.

Both data sets contain models of non-rotating stars of different EoSs, providing the values of a wide range of parameters of these neutron stars. There is some overlap in the parameter space considered within each data set, but both data sets were generated independently as part of different research projects.

While \textbf{Data set A} only covers a subset of the EoSs considered in \textbf{Data set B}, it contains some additional features of non-rotating neutron stars that we can include in our analysis. For a comprehensive discussion of the EoSs covered in each data set we refer to each respective publication. For an overview, we show in Figure~\ref{fig:EoS} the mass-radius relations of a wider range of EoSs, taken from~\cite{Kuan:2021jmk}, of which the EoSs considered in this work are a subset.

The main purpose of utilizing two different data sets is that it allows us to investigate in how far our qualitative observations regarding, e.g., the relative performance of each correlation measure, generalize to different data. To this end, we treat each data set independently, and do not merge the data to obtain one larger data set. By observing the same behavior independently in both data sets increases the confidence that the observations made here also generalize to other data.

\begin{table}[t]
\def\arraystretch{1.5}%
\centering
\begin{tabularx}{0.8\columnwidth}{X | c | c}
Name & Symbol & Data set \\\hline
Gravitational Mass & $\overline{M} = M/M_\odot$ & A, B \\
Radius & $R$ & A, B \\
Square Root of\newline Average Density & $\tilde \rho = \sqrt{M/R^3}$ & A, B \\ 
Compactness & $C = M/R$ & A, B \\
Moment of Inertia & $\bar I = I/M^3$ & A \\
Effective Compactness & $\eta = \sqrt{M^3/I}$ & A \\\hline
$f$-mode frequency & $\omega_f = 2 \pi f_2$  & A, B\\
$g$-mode frequency & $\omega_{g_1} = 2 \pi f_{g_1}$ & B \\
Tidal Deformability & $\bar\lambda = \frac{\lambda}{M^5}$ & A, B \\
\end{tabularx}
\caption{Neutron star features considered in this paper. The last column indicates whether these features are available in Data sets A or B.}
\label{tab:features}
\end{table}

\subsection{Neutron Star Features}
The features considered in our analysis are obtained through either the direct integration of the TOV equations, or through first-order perturbation of the non-rotating neutron star models. The formal description on how these features are obtained are presented in the previous publications that introduced this data~\cite{2021PhRvD.104b3005M,Kuan:2021jmk}. We here summarize the properties of these features. Table~\ref{tab:features} gives an overview of all the features mentioned here.

The first group of features is comprised of macroscopic equilibrium features of the computed neutron star models. In a first step, this includes the \emph{gravitational mass} $M$ (typically normalized $\bar M$ = $M/M_\odot$, where $M_\odot$ is the solar mass), the \emph{radius} $R$ and the \emph{compactness} $C = M/R$. In a second step, we here also consider other neutron star features that have been identified in the literature as useful in the construction of \AU relations. This includes the square-root of the \emph{average density} $\tilde \rho = \sqrt{M/R^3}$, the \emph{moment of inertia} $I$ (typically normalized $\bar I = I/M^3$) and \emph{effective compactness} $\eta = \sqrt{M^3/I} $ of the neutron star.

All of these equilibrium features we try to correlate to various perturbative features that are computed using linear perturbations: this includes the \emph{tidal deformability} $\lambda$ (typically normalized $\bar \lambda = \lambda/M^5$), the \emph{(angular) $f$-mode frequency} $\omega_f$ and the \emph{(angular) $g$-mode frequency} $\omega_{g_1}$ (we here only consider the first $g$-mode frequency for brevity, but keep the given notation to go along with the notation presented in~\cite{Kuan:2021jmk}). To keep in line with a commonly used notion in the literature~\cite{PhysRevLett.77.4134,Andersson98}, we will denote relations involving the latter as \emph{astroseismological} relations.

\section{Bivariate Correlation Analysis}\label{sec:bivariate}
The simplest \AU relations try to directly relate two different features of neutron stars, i.e., they are \emph{bivariate} relations. We believe that by evaluating the correlation between different features, we can automate finding such bivariate relations to a high degree. The main issue, however, is identifying which correlation measure is best suited to the task of finding \AU relations (for neutron stars). 

In this section, we first discuss the concept of \emph{linear} correlation and the corresponding linear correlation measure (Pearson correlation). We then introduce three additional \emph{non-linear} measures of relation that will allow us to find universally related features that are not uncovered by linear correlation. 

\subsection{Linear Correlation}
Throughout this paper, we differentiate between \emph{linearly} and \emph{non-linearly} related features. On a basic level, we will use these two terms to describe the structure that is visually apparent in the scatter plot of a given feature pair: we consider two features to be linearly related if their functional relation can be well approximated by a linear function, and non-linearly related if they show any other kind of functional relation that is not represented by a linear function. 

The degree of linear relation between these features is more formally quantified by their \emph{linear correlation}, otherwise known as Pearson Correlation. Assume that the values of the two considered features are given by two random variables $X$ and $Y$. 
Then the Pearson correlation coefficient $\rho$ of $X$ and $Y$ is given by
\begin{equation}
\rho(X,Y) = \frac{\cov(X,Y)}{\sigma_X \sigma_Y}
\label{eq:pearson}
\end{equation}
where, given the means $\overline X$ and $\overline Y$ of the random variables, $\cov(X,Y)$ is the \emph{covariance} of the two random variables given by
\begin{equation}
\cov(X,Y) = \mathbb{E}\left[\left(X-\overline X\right)\left(Y - \overline Y\right)\right],
\end{equation}
with $\mathbb{E}$ being the expected value of a random variable, and $\sigma_X$ and $\sigma_Y$ their \emph{standard deviations} given by
\begin{equation}
\sigma_X^2 = \mathbb{E}\left[X^2\right] - \mathbb{E}\left[X\right]^2 .
\end{equation}
Note that, by definition, $-1 \leq \rho \leq 1$, and we generally consider the absolute value $\lvert \rho \rvert$ to quantify the degree of linear relation: a value $\lvert \rho \rvert \sim 1$ indicates a perfect linear linear relation between the random variables, whereas $\lvert \rho \rvert \sim 0$ indicates no linear relation.

By considering features pairs with high correlation value $\rho$, the Pearson correlation coefficient can be used to find universally related feature pairs that show a sufficiently strong linear relation. Typically, the threshold for the Pearson correlation coefficient has to be chosen to be high in order to avoid too many false positives, and as such only feature pairs that show a high degree of linear relation will be identified as universally related. The exact choice of this threshold depends highly on the use case, and we will explore this issue further in the following sections.

\subsection{Measures of Non-linear Relation}
We denote the relation of any feature pair that shows some type of universal relation, but that is not classified as linearly related by using the Pearson correlation coefficient, as \emph{non-linear}. Our hypothesis for this work is that by utilizing suitable measures that quantify some type of (not necessarily linear) relation between two random variables, we will be able to identify such non-linearly related features.  

\medskip
\noindent
\textbf{Distance Correlation.}
The first measure that we consider is \emph{Distance correlation} (DistCor), which was specifically introduced as a generalization of Pearson Correlation to identify pairs of random variables that show any kind of linear or non-linear relations. The Distance correlation\cite{DistanceCor} $\dCor$ of two random variables $X$ and $Y$ is defined similarly to the Pearson correlation by 
\begin{equation}
\dCor(X,Y) = \frac{\dCov^2(X,Y)}{\sqrt{\dVar(X)\dVar(Y)}}
\label{eq:distance}
\end{equation}
where, by definition, we have that $0 \leq \dCor \leq 1$, and $\dCor = 0$ if and only if $X$ and $Y$ are statistically independent. 

For Distance correlation, the standard notions of covariance and standard deviation are replaced by sample distance covariance $\dCov$ and distance standard deviation $\dVar$. Similar to covariance and standard deviation, which are computed based the distance of each sample from the means of the random variables, $\dCov$ and $\dVar$ denote quantities that are instead based on the \emph{pairwise} distance of all samples to each other as well as the sample means. As their definitions are slightly longer, and do not necessarily provide critical insight required for the rest of the paper, we refer to the original publication~\cite{DistanceCor} for their full definitions. 

\medskip
\noindent
\textbf{Mutual Information.}
The second measure that we consider is \emph{Mutual Information} (MI), which is an information-theoretic quantity that measures how much we can learn about one random variable $Y$ by having knowledge of another random variable $X$ (or vice versa), and is zero exactly when the two distribution are independent (i.e. knowledge about $X$ does not tell us anything about $Y$). As a quantity, it measures how many bits can be saved if we try to binary encode $Y$ while assuming knowledge of $X$ (in contrast to binary encoding $Y$ on its own without any further knowledge).

The Mutual Information\cite{MutualInf} $I(X;Y)$ of two random variables $X$ and $Y$ is given by
\begin{equation}
I(X;Y) = \sum_{x,y} P_{XY}(x,y) \log \frac{P_{XY}(x,y)}{P_X(x) P_Y(y)}
\label{eq:mutual}
\end{equation}
where $P_{XY}$ is the \emph{joint probability distribution} of $X$ and $Y$ given by
\begin{equation}
P_{XY}(x,y) = Pr[X = x, Y=y],
\end{equation} 
and $P_X$ and $P_Y$ are the marginal distributions given by
\begin{equation}
P_X(x) = \sum_y P_{X,Y}(x,y)\,.
\end{equation}
By definition, we only have that $0 \leq I(X;Y)$, i.e. this measure is not normalized to a range between 0 and 1. While there exist variations that transform Mutual Information into a metric within this domain, we do not strictly require these properties for our use case. As such, we decided to remain with this basic definition. The notion of Maximal Information discussed below presents an extension that is normalized to 1.

Technically, the above definition for Mutual Information is for discrete variables, while our use case is centered around continuous random variables. However, in practice, the data vectors we use are discrete, and computational methods can be used to estimate the continuous sample distribution from the actual, discrete sample vectors. In this paper, we rely on the \textsc{mutual\_info\_regression} method implemented in the \textsc{sklearn} Python package.

\medskip
\noindent
\textbf{Maximal Information.}
The measure of Maximal information (MaxI)~\cite{MaxInf} is a direct extension of Mutual Information to continuous variables. It is based on the binning-based method to estimate the Mutual Information where continuous random variables are discretized by transforming them into histograms of some fixed bin size. Instead of arbitrarily choosing this bin size, leading to varying degrees of accuracy in your Mutual Information estimate, Maximal Information computes the Mutual Information for a series of histograms of varying bin sizes and finally chooses the binning that maximizes the Mutual Information. That is, the maximal information coefficient $\MIC$ of two random variables $X,Y$ is given by
\begin{equation}
\MIC(X;Y) = \max_{B} \frac{I(X;Y)}{\log_2(\min(B_X, B_Y))}
\label{eq:mic}
\end{equation}
where $B$ is the total number of used bins (typically with some upper bound, cf.~\cite{MaxInf}), and $B_X$ and $B_Y$ are the number of bins used for $X$ and $Y$ respectively. By definition, $0 \leq \MIC(X;Y) \leq 1$, where a value of 0 indicates statistical independence of the random variables, whereas a value of 1 indicates a strong relation. To compute the maximal information between two vectors, we will utilize the \textsc{minepy} package for Python~\cite{10.1093/bioinformatics/bts707}.

\subsection{Comparison of Measures}
As mentioned above, the main issue with the more prominent Pearson correlation measure is that it only identifies linearly related features. While we can adjust to this to some degree by computing some function values of our features (i.e. computing some polynomials or exponential function on the features values), this can become fairly cumbersome in practice. In recent years, especially with the advent of Big Data and the necessity of finding non-linear correlations in various applications, the other above mentioned correlation measures have been developed~\cite{DistanceCor,MaxInf}. 
The main idea behind them is that instead of looking for a global, linear correlation, they instead approximate global correlation by finding local (linear correlation), i.e. correlation of data points that are in close proximity, and generalize it over the whole data set. This applies to both Distance correlation, which to some degree generalizes the Pearson correlation in such a manner, and Maximal Information, which directly generalizes the measure of Mutual Information.

A similar comparison has already been performed in the past by Clark~\cite{clark2013comparison}. They find that, in particular for non-linear relations, distance correlation and mutual/maximal information outperform Pearson correlation in identifying correlated variables. Our purpose for this work is to verify that the same observations can also be made for the use case of finding \AU relations in neutron star model data, and evaluate which correlation measure indeed performs best for this use case. 

\subsection{General Methodology}
Our general approach to evaluating the different correlation measures introduced above, and also for later automatically finding bivariate \AU relations, is the following:
\begin{enumerate}
\item Obtain neutron star model data with features $F_1, \ldots, F_n$ from theoretical/numerical computations.
\item Compute the pairwise correlation of all feature pairs using on of the above correlation measures. This provides us with the correlation matrix $\mathbf{M}$, where the entry $\mathbf{M}_{i,j}$ is the correlation between features $F_i$ and $F_j$.
\item Specify a correlation threshold $\tau$ above which we will consider feature pairs correlated, i.e. find all entries in $\mathbf{M}$ with
\begin{equation}
\mathbf{M}_{i,j} \geq \tau\,.
\end{equation}
This threshold will depend on the correlation measure used, and finding the best value for it is something we want to achieve here, but might need to be further explored in future work.
\item For each selected feature pair, choose a suitable model. Here, model denotes the expected functional relation between the two selected features. This can be, e.g., a linear, polynomial, exponential model, etc. Model selection is a notoriously difficult task in data analysis, and we will here simply choose to evaluate a number of preset templates for the functional relations, and choose the one with the best fit after the following step.
\item Fit the model to the given data to determine the coefficients of the best fit for the \AU relation.
\end{enumerate}

\section{Bivariate Universal Relations}\label{sec:bivariate_relations}
In the following, we inspect the \AU relations found by the correlation measures we discussed in the previous section. For each relation, we will also indicate the correlation value obtained by each respective measure. This will allow us to inspect in which cases each of the correlation measures succeed or fail in correctly identifying features that are suited for \AU relations.

Since the features we correlate cover very different ranges of values, we will evaluate the quality of each proposed \AU relation through the \emph{average relative error} $\bar e$ given by
\begin{equation}
\bar e = \frac{1}{n} \sum_i \frac{\lvert \hat y_i - y_i \rvert}{\lvert y_i\rvert}
\end{equation}
where $\hat y_i$ is the value predicted by the \AU relation, and $y_i$ the actual data point.

In some cases, our automated approach will find an exponential relation between two feature that we are analyzing. We find that by instead fitting for the logarithm of the target feature we achieve better universality. In such cases, after performing the correlation analysis on the regular features, we therefore manually fit a polynomial relation between the logarithm of the target feature and the independent feature. Note that the correlation values, however, will still be given between the regular features, and not after applying the logarithm, as this is how the features are fed into the automatic method described in Section~\ref{sec:bivariate}.

As discussed in Section~\ref{sec:data_sets}, we derive relations independently for both data sets to demonstrate to some degree that our approach generalizes across different data. For each relation, we will indicate form which data set it specifically was derived. In most cases, the choice of data set for a given relation was predicated by the features available within each data set (cf. Table~\ref{tab:features}).

A table summarizing all universal relations presented in this section can be found in the Conclusion~\ref{sec:conclusion}.
\subsection{Tidal Deformability Relations}
\begin{figure}
\centering
\includegraphics[width=0.9\columnwidth]{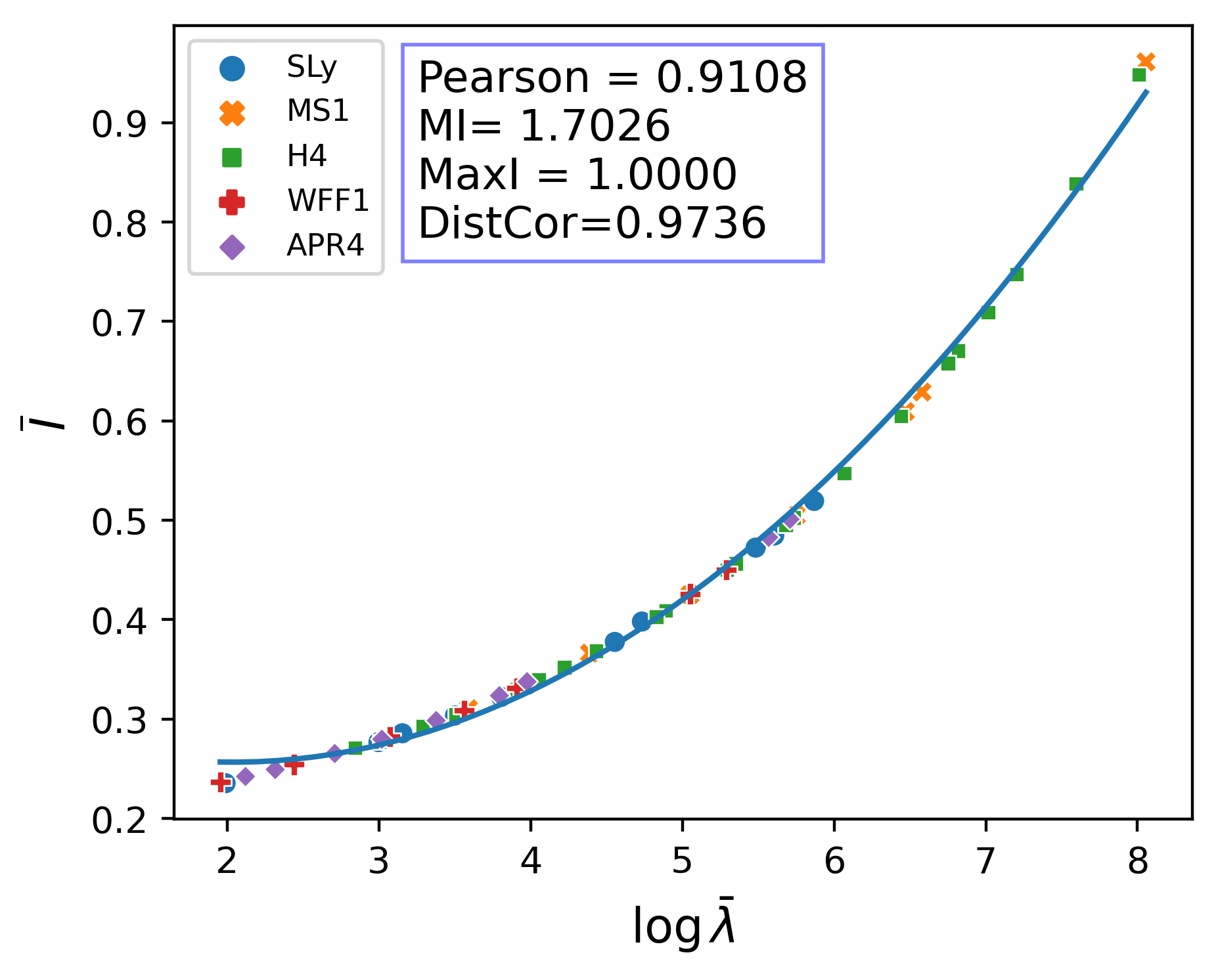}
\caption{Universal relation between the normalized tidal deformability $\bar \lambda$ and normalized moment of inertia $\bar I$, derived from \textbf{Data set A} and as given by Equation~\eqref{eq:bivariate_0} (cf. Yagi and Yunes~\cite{Yagi365}).}
\label{fig:bivariate_0}
\end{figure}
In Figure~\ref{fig:bivariate_0} we show a \AU relation between the normalized tidal deformability $\bar\lambda$ and the normalized moment of inertia $\bar I$, derived from \textbf{Data set A}. This relation was also previously put forward by Yagi and Yunes~\cite{Yagi365} as part of their \emph{I-Love-Q} relations. The best fit for this relation is given by the function 
\begin{equation}
\bar I = \num{0.01860855} \log \bar\lambda^2 - \num{0.07584032}\log \bar\lambda + \num{0.33375708}\,.
\label{eq:bivariate_0}
\end{equation}
This relation achieves an average relative error of $\num{0.020266944213251526}$.

\begin{figure}
\centering
\includegraphics[width=0.9\columnwidth]{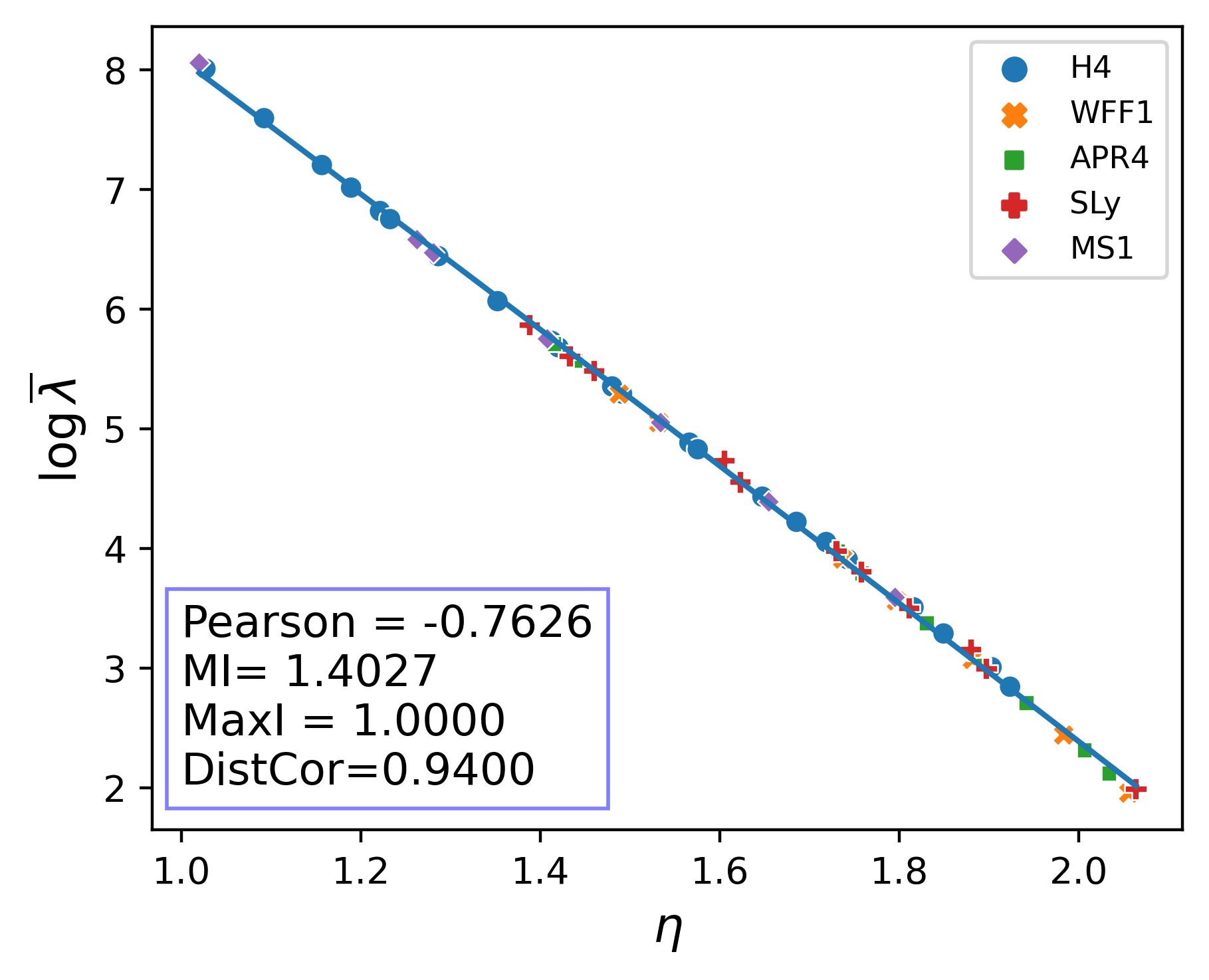}
\caption{Universal relation between normalized tidal deformability $\bar \lambda$ and effective compactness $\eta$, derived from \textbf{Data set A} and as given by Equation~\eqref{eq:bivariate_1}. The correlation values are given for the correlation between $\eta$ and $\bar\lambda$.}
\label{fig:bivariate_1}
\end{figure}
In Figure~\ref{fig:bivariate_1} we show a \AU relation between the effective compactness $\eta$ and the logarithm of the normalized tidal deformability $\log \bar\lambda$, derived from \textbf{Data set A}. A similar relation was also previously proposed by us in the context of a binary neutron star merger connecting the pre-merger binary tidal deformability to the post-merger effective compactness~\cite{2021PhRvD.104b3005M}. 

This is a case in which the automated approach yields an exponential relation between $\eta$ and $\bar\lambda$, and as discussed above, we manually fit a polynomial relation for $\log \bar \lambda$, yielding the relation 
\begin{equation}
\log \bar \lambda = \num{-0.09253471}\eta^2 -\num{5.42472023} \eta +\num{13.60418765}
\label{eq:bivariate_1}
\end{equation}
This relation achieves an an average relative error of $\num{0.007502394271055736}$. In this case, the originally exponential relation between the two features causes the Pearson correlation coefficient in particular to give a comparatively low correlation value of $-\num{0.7626}$. In comparison, the other correlation measures still assign a fairly high correlation value.

\begin{figure}
\centering
\includegraphics[width=0.9\columnwidth]{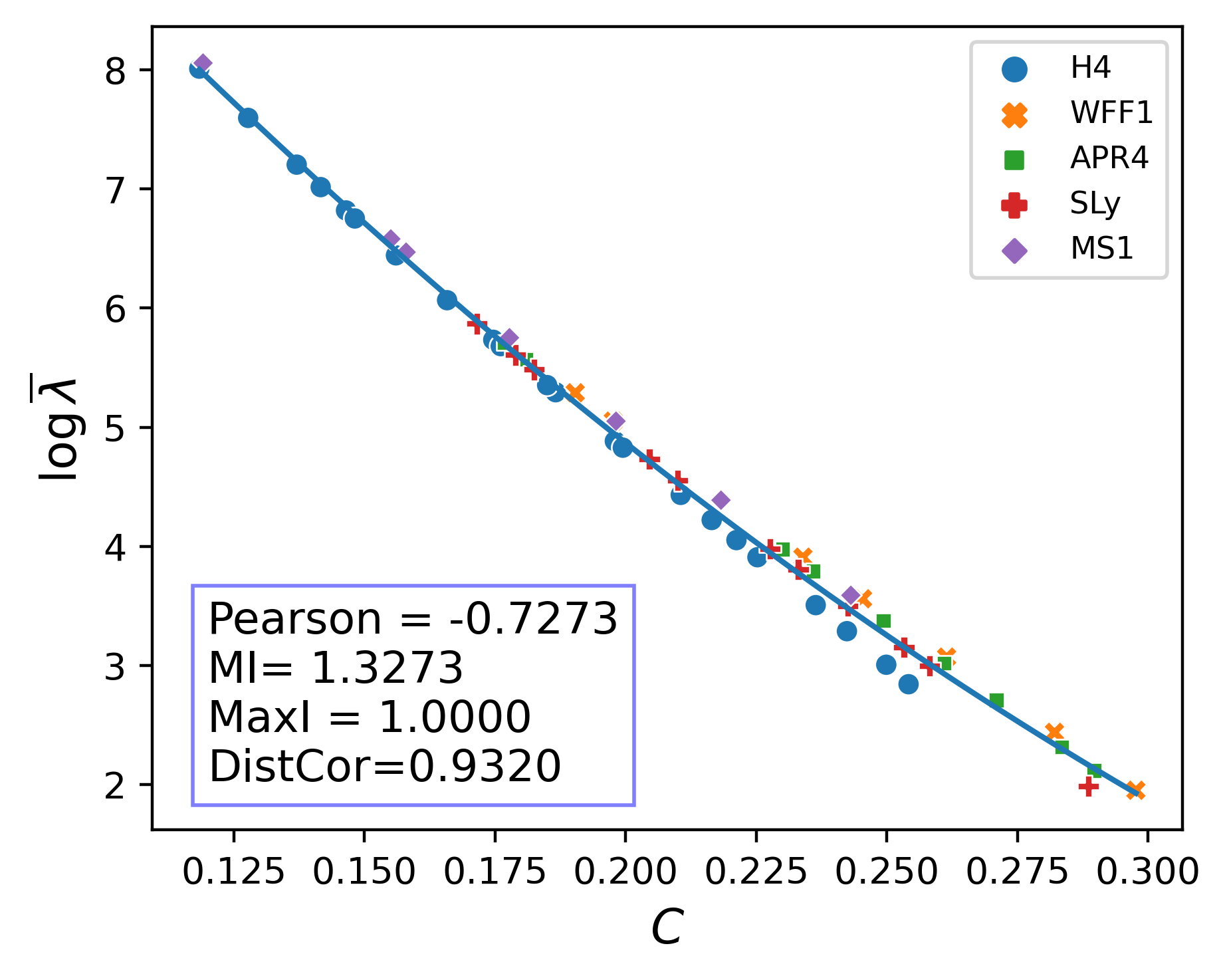}
\caption{Universal relation between the logarithm of the tidal deformability $\log\bar\lambda$ and compactness $C$, derived from \textbf{Data set A} and as given by Equation~\eqref{eq:bivariate_2}. The correlation values are given for the correlation between $C$ and $\bar\lambda$.}
\label{fig:bivariate_2}
\end{figure}
In Figure~\ref{fig:bivariate_2} we show a \AU relation between the compactness $C$ and the logarithm of the normalized tidal deformability $\log\bar\lambda$, derived from \textbf{Data set A}. Such a relation was put forward previously by Jiang and Yagi~\cite{Jiang_2020}, and follows directly from the definition of $\bar\lambda$ in terms of the tidal Love number $k_2$, i.e.
\begin{equation}
\bar\lambda = \frac{\lambda}{M^5} = \frac{2}{3} k_2 \frac{R^5}{M ^5} =\frac{2}{3} k_2 C^5\,.
\end{equation}
The automatic approach again finds an exponential relation between the features $C$ and $\bar\lambda$, and as before, we find that fitting for $\log\bar\lambda$ instead yields the more accurate, universal relations. The manual fit yields the relation 
\begin{equation}
\log \bar \lambda =  \num{46.1226271} C^2 -\num{53.04505671} C + \num{13.63341181}
\label{eq:bivariate_2}
\end{equation}
This relation achieves a relative error of \num{0.019736506644986718}. As before, the regular features have a non-linear, exponential relation for which the Pearson correlation measure again assigns a comparatively low correlation value of \num{-0.7273}, while the other correlation measures still assign high correlation values.

\subsection{Astroseismological Relations}\label{sec:bivariate_osci}
We here present some of the astroseismological, universal relations we were able to find for the $f$-mode and $g$-mode oscillation frequencies.

\begin{figure}
\centering
\includegraphics[width=0.9\columnwidth]{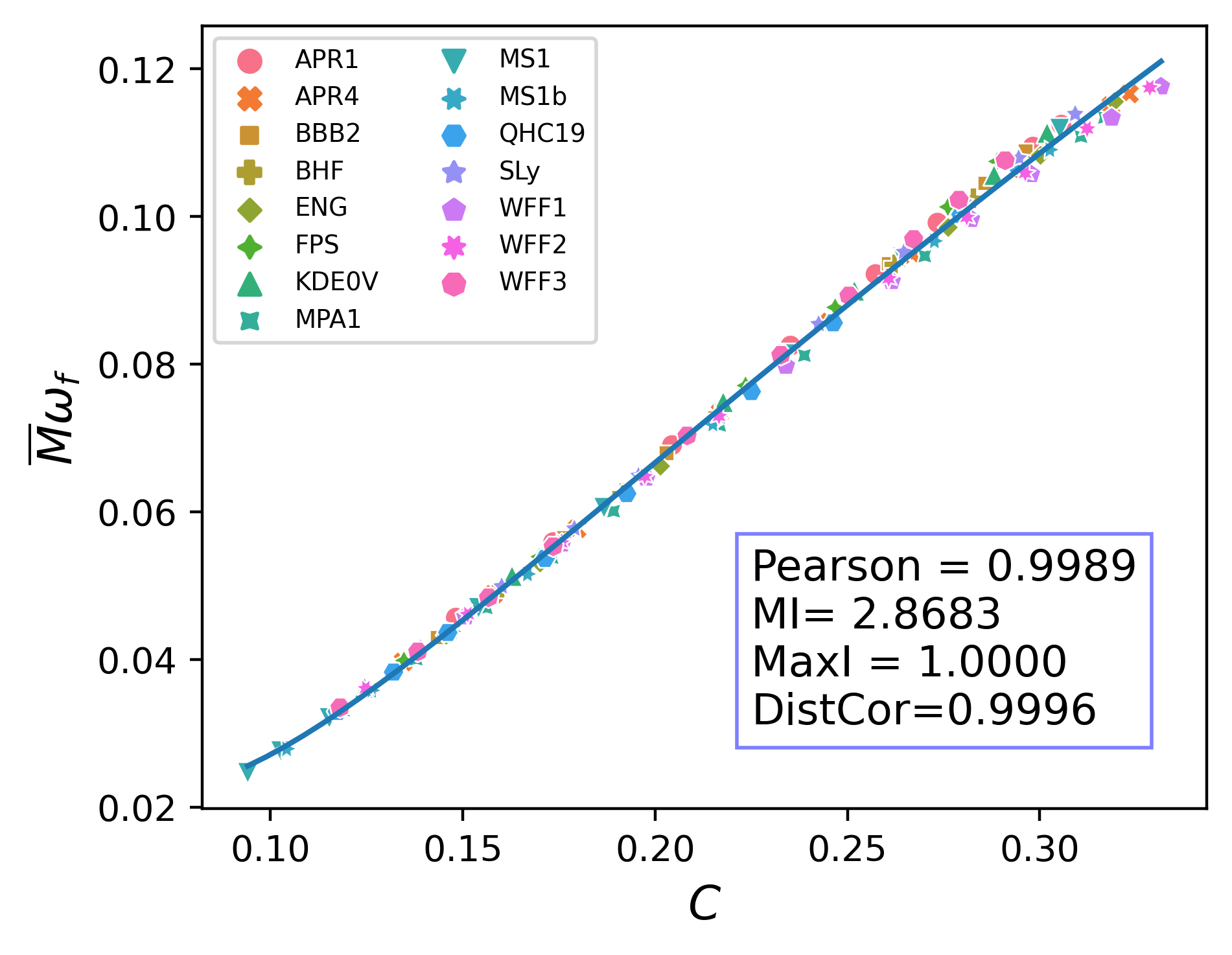}
\caption{Astroseismological relation between the normalized $f$-mode frequency $\bar M\omega_f$ and the compactness $C$, derived from \textbf{Data set B} and as given by Equation~\eqref{eq:bivariate2_0} (cf. Tsui and Leung~\cite{PhysRevLett.95.151101}).}
\label{fig:bivariate2_0}
\end{figure}
In Figure~\ref{fig:bivariate2_0} we show a \AU relation between the compactness $C$ and the normalized $f$-mode frequency $\bar M \omega_f$, derived from \textbf{Data set B}. This relation was previously put forward by Tsui and Leung~\cite{PhysRevLett.95.151101}. The best fit for this relation is given by the function
\begin{equation}
\bar M \omega_f = \num{0.04212737} \log C^2+ \num{0.22183278} \log C + \num{0.3145776}\,.
\label{eq:bivariate2_0}
\end{equation}
This relation achieves an average relative error of \num{0.01148544725114905}. While the optimal fit is given by a logarithmic relation, visually the relation can still be fit fairly well by a linear function. As expected, in this case even the Pearson correlation coefficient assigns a high value of \num{0.9989}, and the other correlation measures also identify a strong relation between these two features.

\begin{figure}
\centering
\includegraphics[width=0.9\columnwidth]{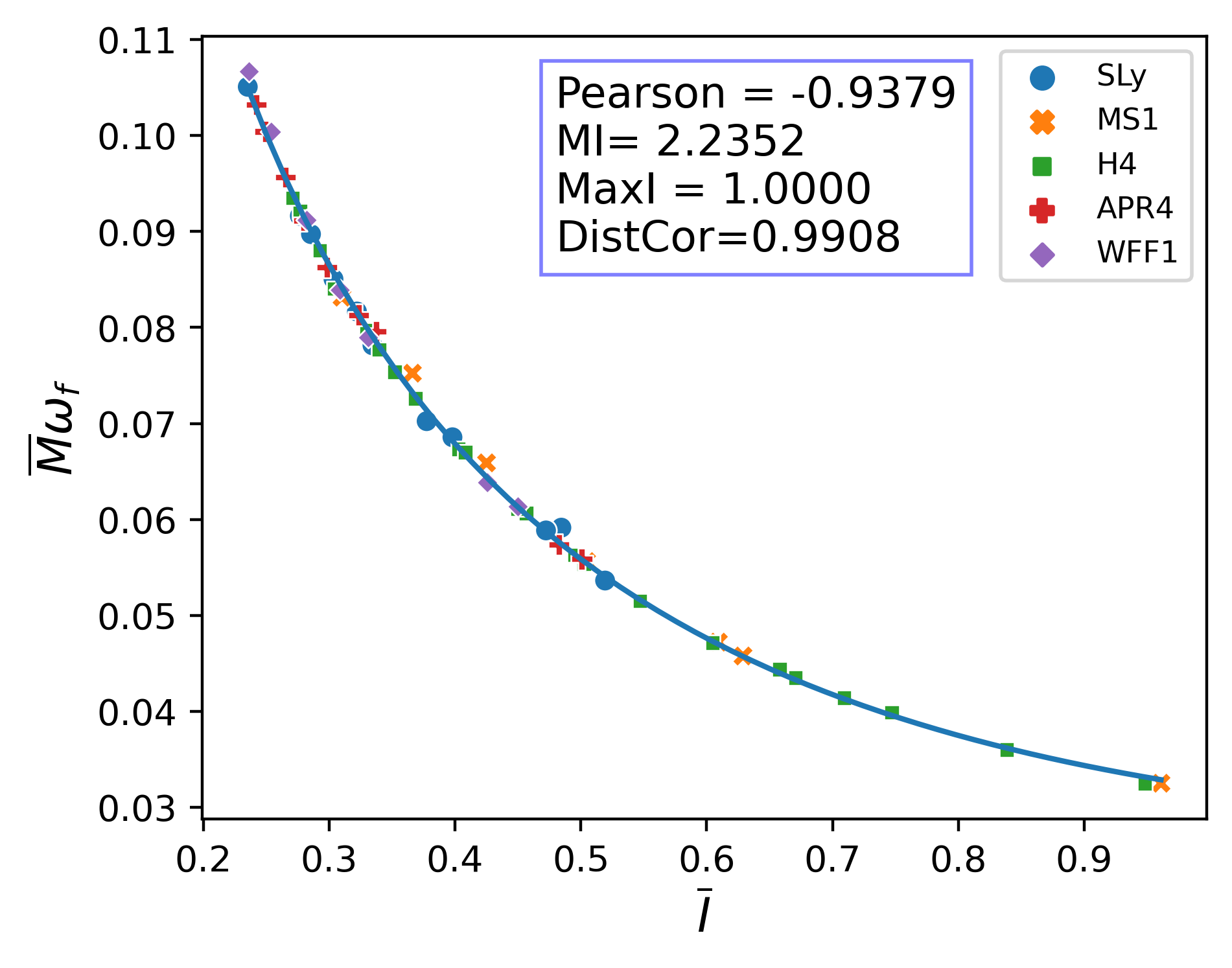}
\caption{Astroseismological relation between the normalized $f$-mode frequency $\bar M\omega_f$ and the normalized moment of inertia $\bar I$, derived from \textbf{Data set A} and as given by Equation~\eqref{eq:bivariate2_1}.}
\label{fig:bivariate2_1}
\end{figure}
In Figure~\ref{fig:bivariate2_1} we show a \AU relation between the normalized moment of inertia $\bar I$ and the normalized $f$-mode frequency $\bar M \omega_f$, derived from \textbf{Data set A}. This relation follows straight-forwardly by combining the relation by Tsui and Leung~\cite{10.1111/j.1365-2966.2005.08710.x} between the $f$-mode and compactness $C$, with the understanding that the compactness $C$ and effective compactness $\eta$ can often be used interchangeably in such general relativistic relations. However, to our knowledge, this is the first time that this relation is presented explicitly. 

The best fit for this relation is given by the function
\begin{equation}
\bar M \omega_f = \num{0.02130106} \log {\bar I}^2 - \num{0.01956669} \log \bar I + \num{0.03207015}\,.
\label{eq:bivariate2_1}
\end{equation}
This relation achieves an average relative error of \num{0.006879103949206297}. The best fit is given by a logarithmic relation, and visually the relation also does not seem to allow a good fit by a linear function. Still, in this case, the Pearson correlation still applies a comparatively high value of \num{-0.9379}. The remaining measures also identify a strong relation between the features.

\begin{figure}
\centering
\includegraphics[width=0.9\columnwidth]{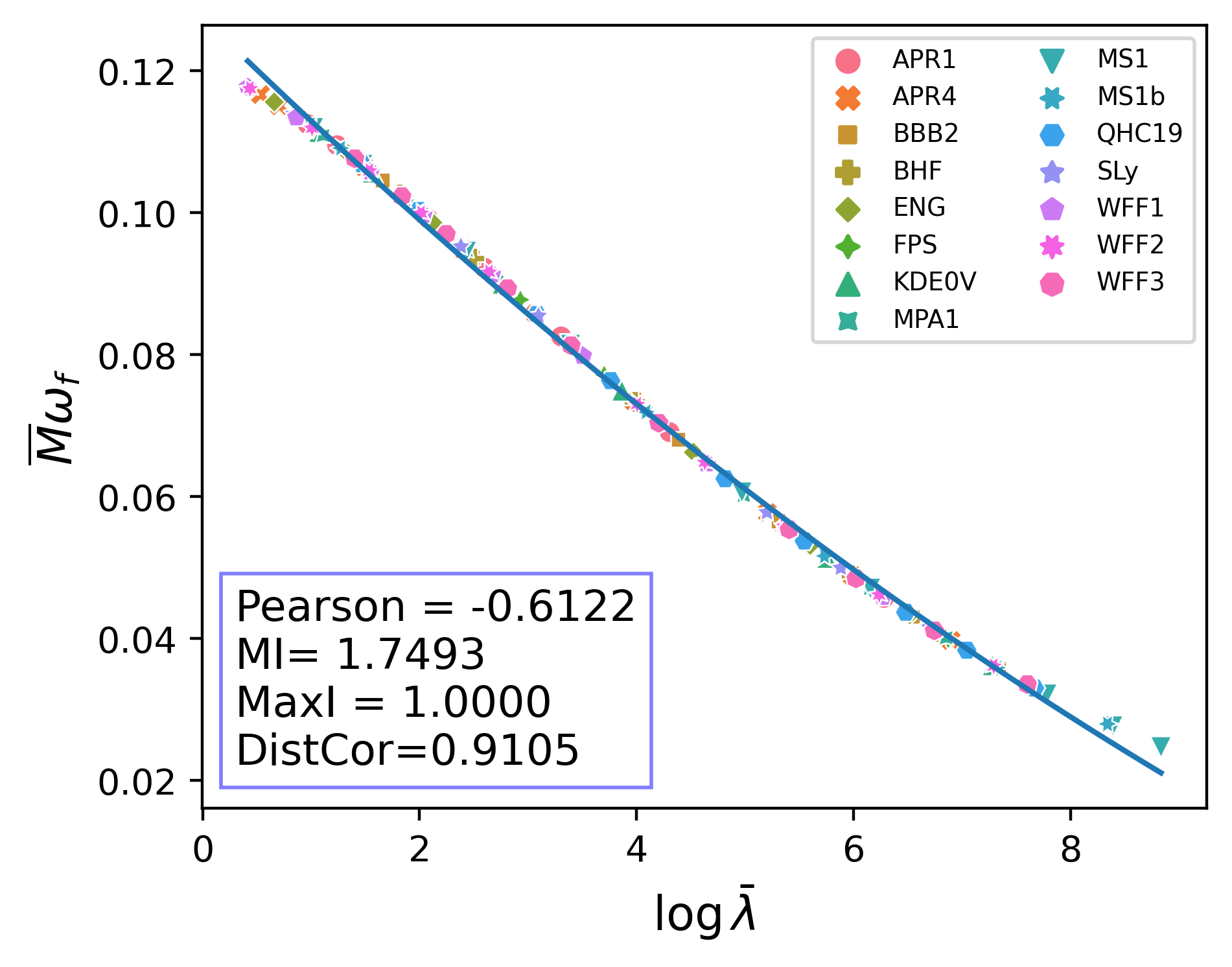}
\caption{Astroseismological relation between the normalized $f$-mode frequency $\bar M\omega_f$ and normalized tidal deformability $\bar \lambda$, derived from \textbf{Data set B} and as given by Equation~\eqref{eq:bivariate2_2} (cf. Chan et al.~\cite{chan2014multipolar}).}
\label{fig:bivariate2_2}
\end{figure}
Figure~\ref{fig:bivariate2_2} shows a \AU relation between the normalized tidal deformability $\bar \lambda$ and the normalized $f$-mode frequency $\bar M \omega_f$, derived from \textbf{Data set B}. This relation was also previously put forward by Chan et al.~\cite{chan2014multipolar}. The best fit for this relation is given by the function
\begin{equation}
\bar M \omega_f = 0.0003 \log {\bar \lambda}^2 - \num{0.01490679} \log \bar \lambda + \num{0.12746998}\,.
\label{eq:bivariate2_2}
\end{equation}
This relation achieves an average relative error of of \num{0.01425718984920659}. The highly non-linear, logarithmic relation between these features causes the Pearson correlation coefficient to fail to detect the correlation between these features, assigning a value of \num{-0.6122} and even the Distance Correlation assigns a comparatively small correlation value of \num{0.9105}, compared to the previous relations. 

\begin{figure}
\centering
\includegraphics[width=0.9\columnwidth]{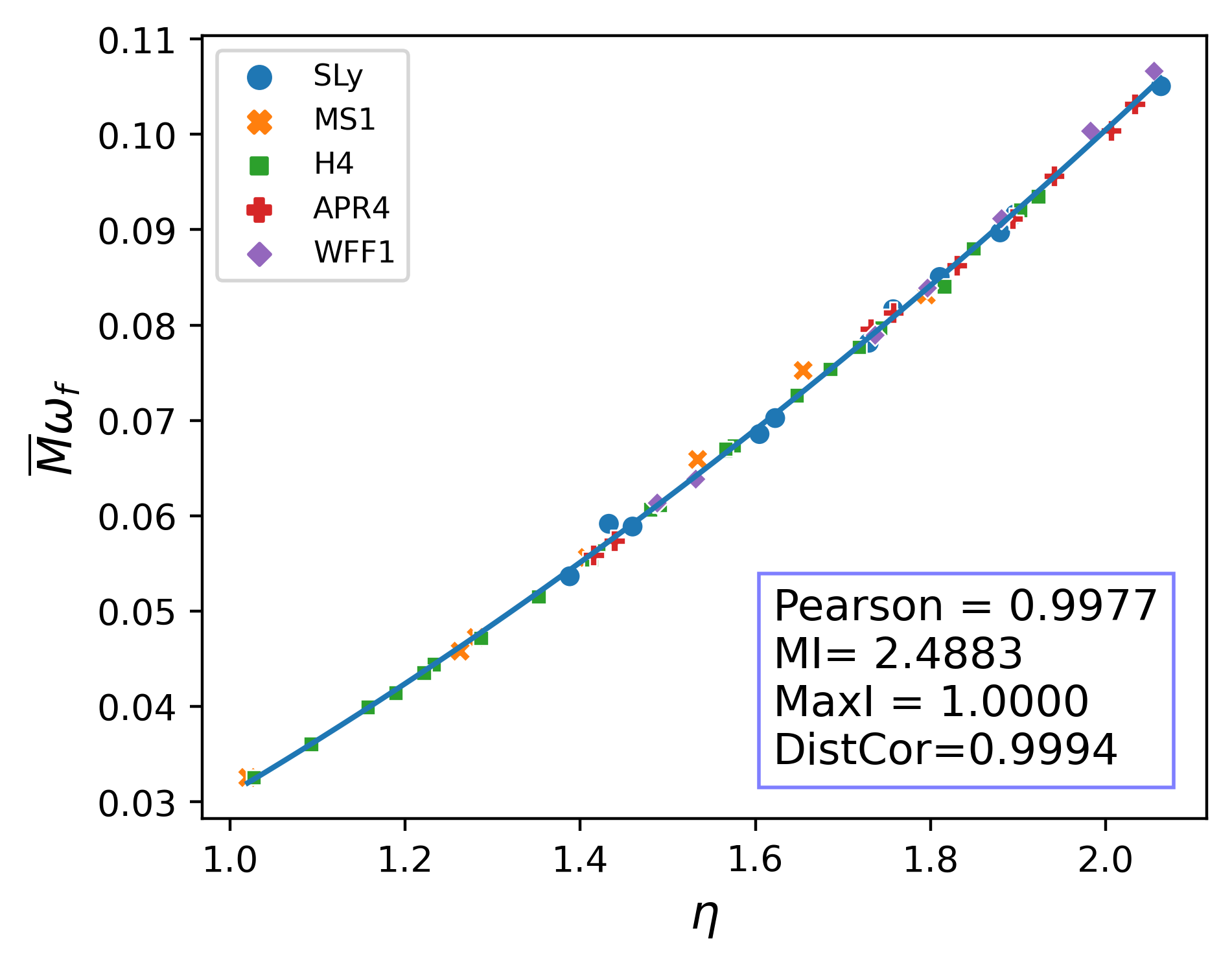}
\caption{Astroseismological relation between the normalized $f$-mode frequency $\bar M\omega_f$ and the effective compactness $\eta$, derived from \textbf{Data set A} and as given by Equation~\eqref{eq:bivariate2_3} (cf. Lau et al.~\cite{lau2010inferring} and Kr\"uger and Kokkotas~\cite{kruger2019fast}).}
\label{fig:bivariate2_3}
\end{figure}
Figure~\ref{fig:bivariate2_3} shows a \AU relation between the effective compactness $\eta$ and the normalized $f$-mode frequency $\bar M \omega_f$, derived from \textbf{Data set A}. This relation was also previously put forward in~\cite{lau2010inferring,PhysRevD.91.044034,kruger2019fast}. The best fit for this relation is given by the function
\begin{equation}
\bar M \omega_f = \num{0.0149526} \eta^2 + \num{0.02475} \eta - \num{0.00885211}\,.
\label{eq:bivariate2_3}
\end{equation}
This relation achieves an average relative error of \num{0.006629639167064198}. Visually, this relation again appears to also allow a good fit through a linear function, which is reflected by all correlation measures (including Pearson Correlation) assigning a high correlation value.

\begin{figure}
\centering
\includegraphics[width=0.9\columnwidth]{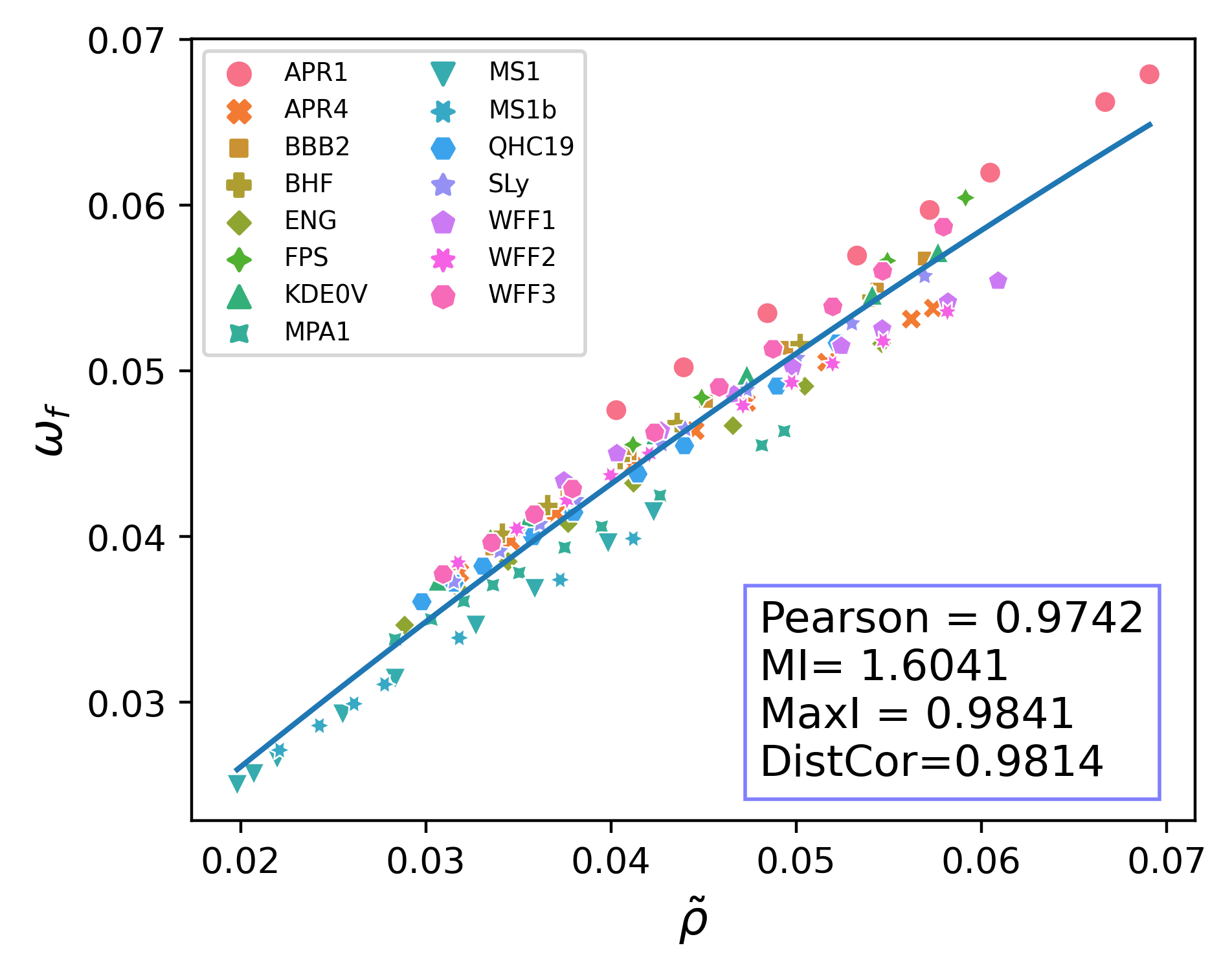}
\caption{Astroseismological relation between the $f$-mode frequency $\omega_f$ and the square-root of the average density $\tilde\rho$, derived from \textbf{Data set B} and as given by Equation~\eqref{eq:bivariate3_0} (cf. Andersson and Kokkotas~\cite{PhysRevLett.77.4134,Andersson98}).}
\label{fig:bivariate3_0}
\end{figure}
Figure~\ref{fig:bivariate3_0} shows a \AU relation between the average density $\tilde\rho$ and the $f$-mode frequency $\omega_f$, derived from \textbf{Data set B}. This relation was also previously put forward by Andersson and Kokkotas~\cite{PhysRevLett.77.4134,Andersson98} and Benhar et al.~\cite{PhysRevD.70.124015}. The best fit for this relation is given by the function
\begin{equation}
\omega_f = \num{-2.19908806} \tilde\rho^2 + \num{0.98518922} \tilde\rho + \num{0.007272}\,.
\label{eq:bivariate3_0}
\end{equation}
This relation achieves an average relative error of \num{0.03499534511917183}. Again, the fact that this relation appears to be mostly linear is reflected in the fact that all correlation measures assign a fairly high correlation value to these two features.

\begin{figure}
\centering
\includegraphics[width=0.9\columnwidth]{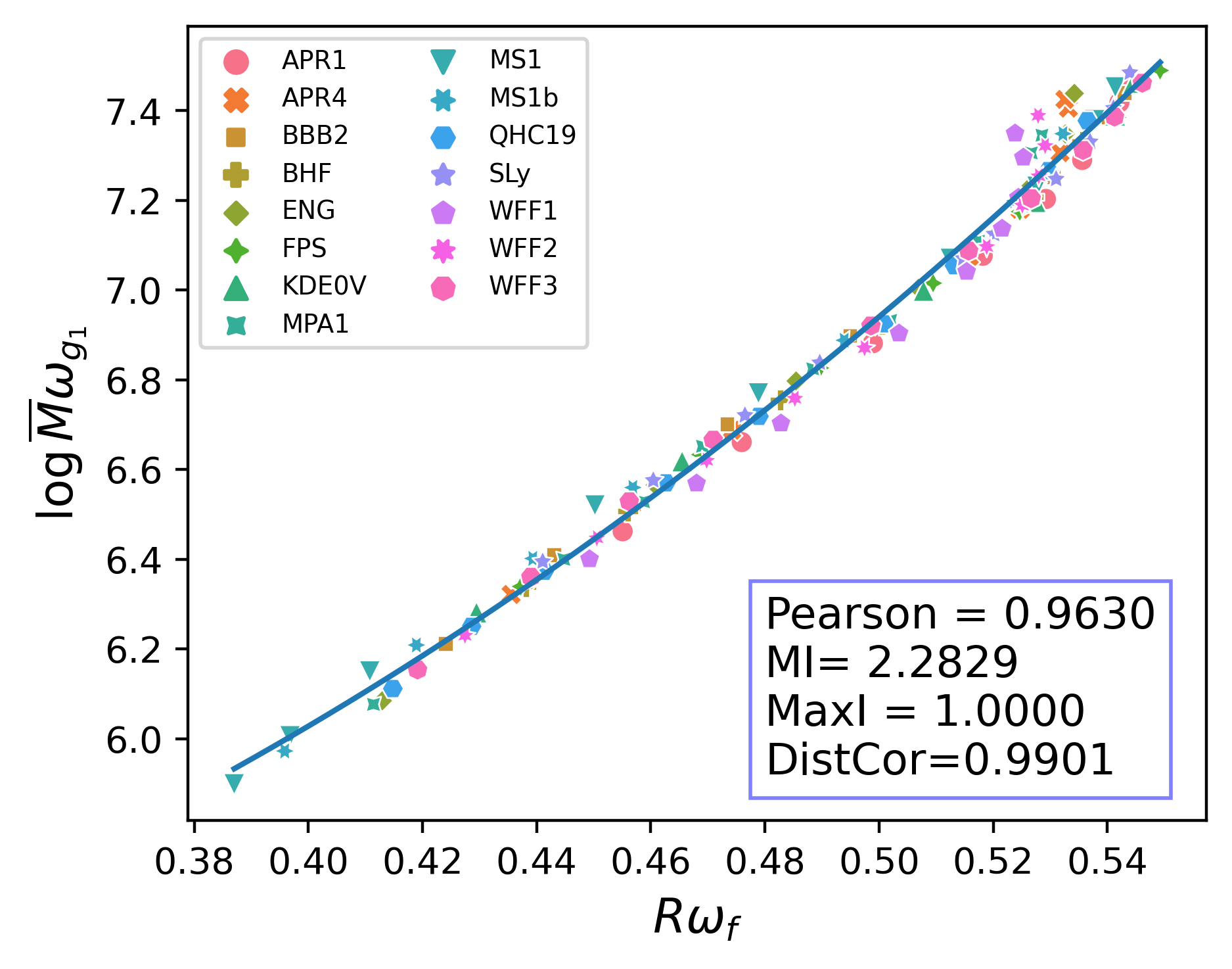}
\caption{Astroseismological relation between the normalized $f$-mode frequency $R \omega_f$ and the logarithm of the normalized $g$-mode frequency $\log \bar M \omega_{g_1}$, derived from \textbf{Data set B} and as given by Equation~\eqref{eq:bivariate2_5} (cf. Kuan et al.~\cite{Kuan:2021jmk}). The correlation values are given for the correlation between $R \omega_f$ and $\bar M \omega_{g_1}$.}
\label{fig:bivariate2_5}
\end{figure}
Figure~\ref{fig:bivariate2_5} shows a \AU relation between the normalized $f$-mode frequency $R \omega_f$ and the logarithm of the normalized $g$-mode frequency $\log \bar M\omega_{g_1}$, derived from \textbf{Data set B}. This relation was also previously put forward by Kuan et al.~\cite{Kuan:2021jmk}. As was the case for the relations in Equations~\eqref{eq:bivariate_1} and~\eqref{eq:bivariate_2}, the automatic method finds an exponential relation between the features $R \omega_f$ and $\bar M\omega_{g_1}$. As before, we find that manually fitting the relation for the logarithm $\log \bar M \omega_{g_1}$ gives the more accurate relation, yielding
\begin{equation}
\log \bar M\omega_{g_1} = \num{16.05167444} \left(R \omega_f\right)^2 -\num{5.32343308} R \omega_f+ \num{5.58908468}
\label{eq:bivariate2_5}
\end{equation}
This relation achieves an average relative error of \num{0.004346355091302037}. Even though the automatic method finds an exponential function to be the best fit between the original features, visually it is apparent that the relation could, to some degree, also be fit by a linear function. As such, even the Pearson correlation coefficient achieves a fairly high correlation value, however notably lower than the other correlation measures.

\subsection{Quantitative Comparison of Correlation Measures}
\begin{figure*}[ht]
\centering
\subfloat[][Precision/Recall curves for \textbf{Data set A}]{\includegraphics[width=0.5\textwidth]{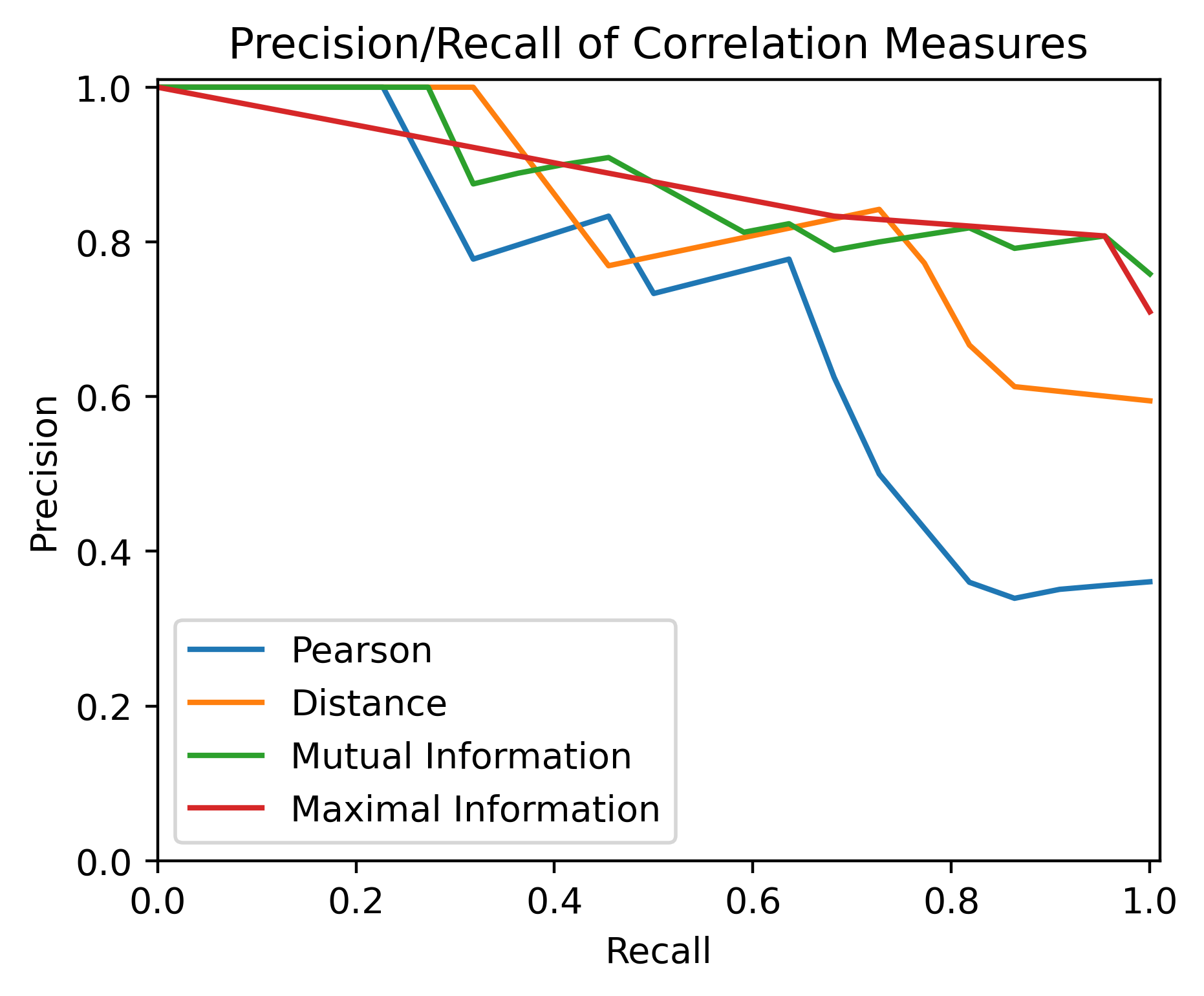}\label{fig:precrec_all_PM}}
\subfloat[][Precision/Recall curves for \textbf{Data set B}]{\includegraphics[width=0.5\textwidth]{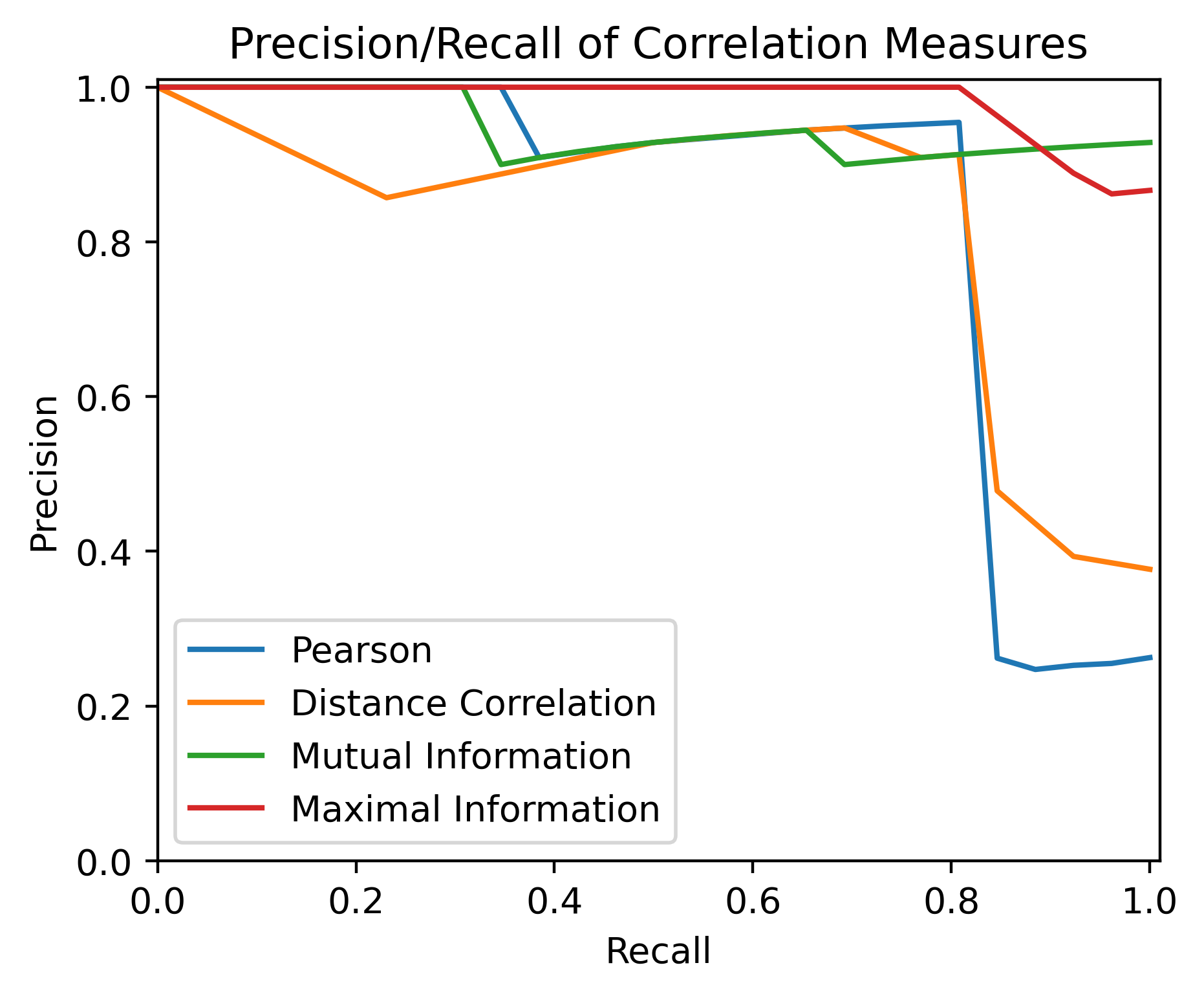}\label{fig:precrec_all}}
\caption{Precision/Recall curves for each correlation measure.}
\label{fig:prec_rec}
\end{figure*}

We can perform a more quantitative analysis and comparison of the four different correlation measures by considering some specific performance measures commonly used in statistics. To define these, we first introduce a few notions for binary classifiers. We define them here in terms of our use case of identifying universally related neutron star features: A \emph{true positive} is a pair of features that is universally related, and also identified as such by a given correlation measure. The number of true positives is denoted by $\TP$.

A \emph{false positive} is a pair of features that is \emph{not} universally related, but classified as such. The number of false positives is denoted by $\FP$.

A \emph{false negative} is a pair of features that is universally related, but not classified as such. The number of false negatives is denoted by $\FN$.

A \emph{true negative} is a pair of features that is not universally related, and also not classified as such. The number number of true negatives is denoted by $\TN$.

\begin{figure*}[t]
\centering
\subfloat[][ROC Curves for \textbf{Data set A}]{\includegraphics[width=0.5\textwidth]{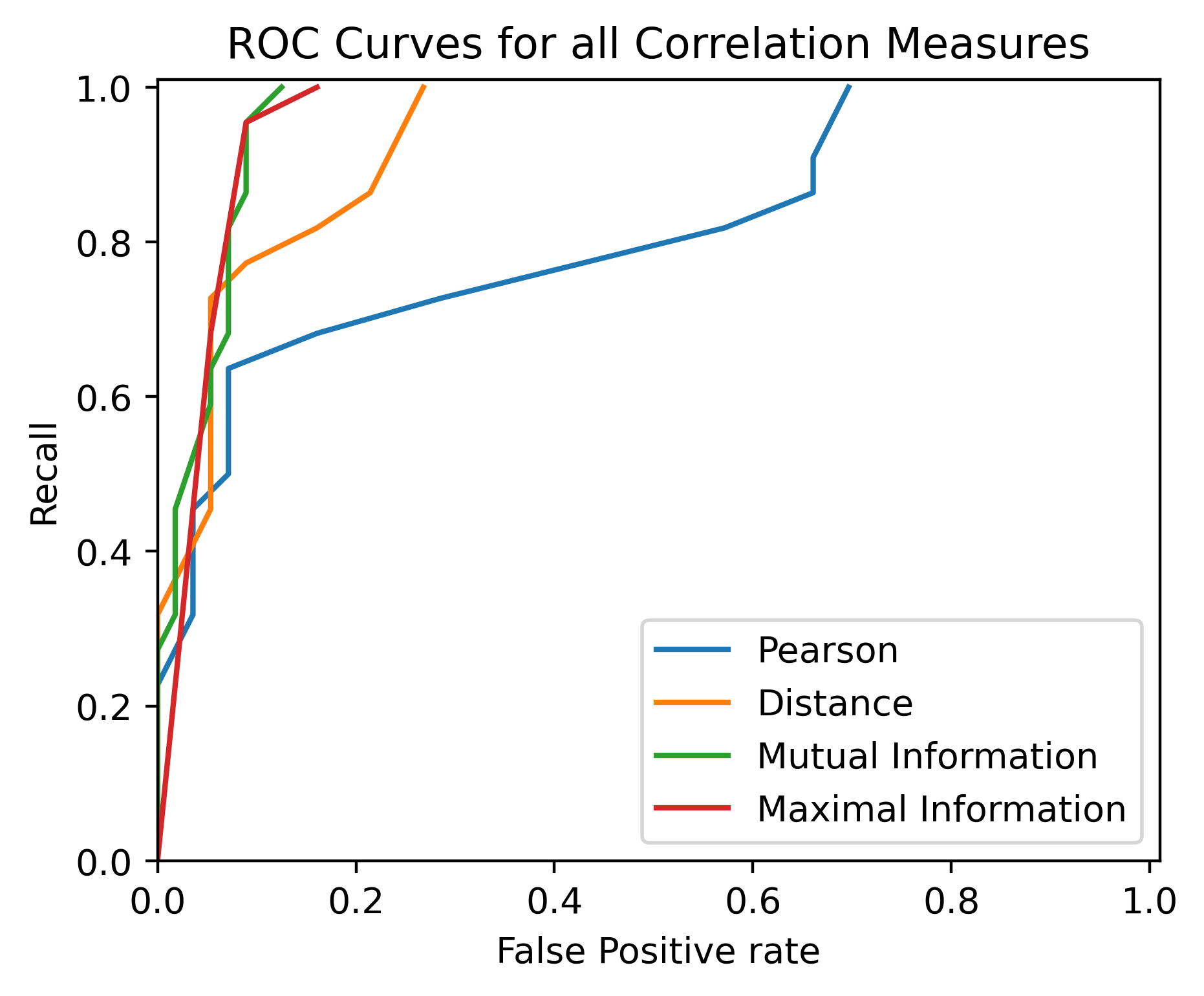}\label{fig:roc_all_PM}}
\subfloat[][ROC Curves for \textbf{Data set B}]{\includegraphics[width=0.5\textwidth]{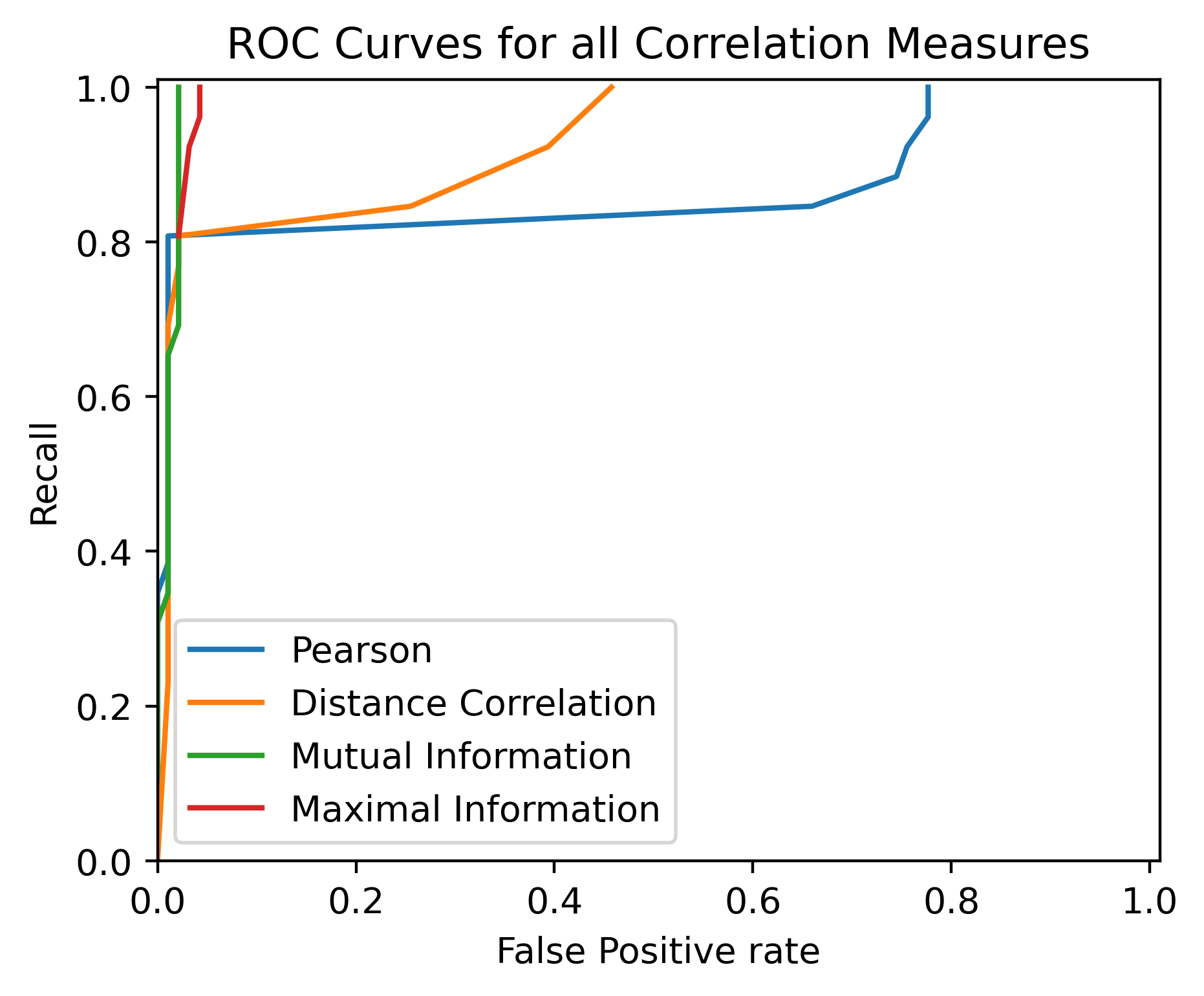}\label{fig:roc_all}}
\caption{ROC (Receiver operating characteristic) curves for each correlation measure.}
\label{fig:roc}
\end{figure*}
Given these notions, we can now define performance measures that quantify how well our classifiers correctly label pairs of features. \emph{Recall}, or \emph{true positive rate} $\TPR$ is the rate at which the classifier correctly labels universally related pairs of features as universally related. It is given by
\begin{equation}
\TPR = \frac{\TP}{\TP + \FN}\,.
\end{equation}

\emph{Precision}, or \emph{positive predictive value} ($\PPV$), is the rate of pairs of features classified as universally related that are in fact universally related. It is given by
\begin{equation}
\PPV = \frac{\TP}{\TP + \FP}\,.
\end{equation}

Finally, the \emph{fallout}, or \emph{false positive rate} ($\FPR$), is the rate at which not related pairs of features are classified as being universally related. It is given by
\begin{equation}
\FPR = \frac{\FP}{\FP + \TN}\,.
\end{equation}

We can now compute the precision, recall and fallout for each correlation measure at a given classification threshold $\tau$, and compare how these performance measures develop with $\tau$. Ideally, we would like to achieve high recall, while keeping precision high, and fallout low.

Typically, one considers the Precision/Recall and ROC-curves for a better understanding on how these quantities evolve with each other. The Precision/Recall curves plot the maximum precision achieved by a classifier for a required recall, and allow us to understand how accurate a positive prediction (i.e. classification as universal relation) is, given a specific correlation measure and classification threshold. We show the Precision/Recall curves for each correlation measure, and one combined plot, in Figure~\ref{fig:prec_rec}.  

The ROC (or receiver operating characteristic) curve plots the recall against the fallout of the classifier. This plot allows us to better understand how many incorrectly classified universal relations we should expect for a given recall requirement. The ROC curves for each correlation measure applied to each of the two data sets considered in this work can be found in Figure~\ref{fig:roc}. 

As we can see in all Figures, the standard Pearson correlation measure is outperformed by the other metrics significantly for most of the recall range. The distance correlation measures, in turn, is also outperformed by the mutual information based measures. Both the kernel-based mutual information measure, and the maximal information measure show high precision and low fallout for high recall values, identifying them as the preferred measures for the task of identifying universally related features.

Note that the above analyses were performed by manually labeling all feature pairs in our limited data set as either being universally related or not in order to obtain the true/false positive/negative counts. As such, the exact values for each performance measure will most likely vary with different datasets and labels. However, the difference in behavior of each correlation measure appears to be significant enough to warrant the conclusion drawn above.

\medskip
\noindent
\textbf{Ideal Classification Thresholds.}
Closer inspection of both the Precision/Recall and the ROC curves can also provide hints for an \emph{ideal} classification threshold that optimizes the trade-off between recall and precision/false positive rate. This can be done by, e.g., considering thresholds at which precision or false positive rate show drastic changes. This threshold will, however, depend on whether one prioritizes higher recall over precision/false positive rates, or vice versa. For our use case of finding universally related features, we would naturally prefer achieving a higher recall value as false positives can usually be discarded after a simple visual inspection of the corresponding plots. As an example, in Tables~\ref{tab:rec_prec_A} and~\ref{tab:rec_prec_B}, we list the classification thresholds manually selected from the data points of the respective Precision/Recall and ROC curves.
\begin{table}[t]
\def\arraystretch{1.5}%
\centering
\setlength{\tabcolsep}{8pt}
\begin{tabular}{r | c | c | c | c}
Measure & $\tau$ & Recall & Precision & FPR \\\hline
Pearson & \num{0.909091} & \num{0.681818}  & \num{0.625000} & \num{0.160714}\\
DistCor & \num{0.939394} &  \num{0.863636} & \num{0.612903}& \num{0.214286}\\
MI & \num{1.187879}  & \num{0.954545} & \num{0.807692} & \num{0.089286}\\
MaxI & \num{0.959596}  & \num{0.954545} & \num{0.807692} & \num{0.089286}
\end{tabular}
\caption{Classification thresholds $\tau$ and corresponding performance measures derived from the Precision/Recall (cf. Figure~\ref{fig:precrec_all_PM}) and ROC curves (cf. Figure~\ref{fig:roc_all_PM}) for \textbf{Data set A}.}
\label{tab:rec_prec_A}
\end{table}

\begin{table}[t]
\def\arraystretch{1.5}%
\centering
\setlength{\tabcolsep}{8pt}
\begin{tabular}{r | c | c | c | c}
Measure & $\tau$ & Recall & Precision & FPR \\\hline
Pearson & \num{0.939394} & \num{0.807692}  & \num{0.954545} & \num{0.010638}\\
DistCor & \num{0.959596} &  \num{0.807692} & \num{0.913043}& \num{0.021277}\\
MI & \num{1.333333}  & \num{1.000000} & \num{0.928571} & \num{0.021277}\\
MaxI & \num{0.949495}  & \num{1.000000} & \num{0.866667} & \num{0.042553}
\end{tabular}
\caption{Classification thresholds $\tau$ and corresponding performance measures derived from the Precision/Recall (cf. Figure~\ref{fig:precrec_all}) and ROC curves (cf. Figure~\ref{fig:roc_all}) for \textbf{Data set B}.}
\label{tab:rec_prec_B}
\end{table}

Determining a generally optimal classification threshold, however, will require further detailed analysis that is out of the scope of this work as the optimal threshold can often depend on the exact application and data that is used.

\section{Multivariate Correlation Analysis}\label{sec:multivariate}
Until now, we have only considered the functional relation between two features, and tried to find such pairs of features that allow for  \AU  relations across different equations of state. A straightforward extension then, of course, is to look for multivariate relations, i.e. such relations where one predicted/target features is described in terms of a function that depends on more than one explanatory feature.

The field of high-dimensional data analysis is a widely studied field that in particular gained a lot of notoriety in recent times due to the advent of the big data paradigm. While many different approaches, theories and methods have been developed to deal with high dimensionality in data, we will here consider one very prominent method: principal component analysis (PCA). PCA is a dimensionality reduction and feature extraction technique that has been used to great success in various data analysis use cases~\cite{barnett1987origins}. Recently, Soldateschi et al.~\cite{2021A&A...654A.162S} utilized PCA to construct multivariate  \AU  relations for magnetized neutron stars. Here, we will investigate how we can apply PCA in general to identify potential  \AU  relations, and evaluate how well this approach performs on our own data.

\subsection{Finding Multivariate Correlation using PCA}
The main purpose of PCA is to identify the principal directions in which a given data set varies the most: the \emph{principal components} $\mathbf{A}$ of a data set consisting of a set of features $\mathbf{F}$ (i.e. the dimensions of the data set's underlying vector space) are a sequence of vectors (called principal component) $\mathbf{A_i}$ in the space span by $\mathbf{F}$ which
\begin{enumerate}
\item are orthogonal to all previous principal components $\mathbf{A_0},\ldots,\mathbf{A_{i-1}}$.
\item show in the direction of the line that best fits the data set (using least squares regression).
\end{enumerate} 
As a consequence, the principal component $A_0$ shows the direction that maximizes the variance within the data set, while each subsequent principal component covers less and less variance of the data set. One can then choose the $m$ first principal components as the basis vectors of a lower dimensional vector space into which the data set can be projected while retaining most of the variance (i.e. \emph{information}) within the data set.

While the general use case of PCA does not directly match our goal of constructing  \AU  relations, we can make use of the properties of the principal components to potentially find multivariate  \AU  relations. Note that each principal component $\mathbf{A_i}$ represents a linear combination of the features $\mathbf{F}$ that assigning a weight $a_{i,j}$ to each feature $F_j$, which, in this context, is also called the \emph{loading} of $F_j$ in $\mathbf{A_i}$. After computing all principal components, we try to identify those that have a proportionally large loading for our target feature $F$, if any such component exists: usually, if there are no strong correlations within our data that lead to a large variance for $F$, all principal components will show a comparatively small loading for $F$. However, in the case of a principal component that has a large loading for $F$, we might be able to leverage it to construct a  \AU  relation. By projecting the considered features onto the identified principal component and solving for $F$, we potentially obtain a first-order multivariate \AU relations.

Soldateschi et al.~\cite{2021A&A...654A.162S} claim that such universal relations should be found using the last principal component. In the following we will also investigate if this claim is true.

\subsection{General Methodology}\label{sec:multivariate_methodlogy}
We now describe the general methodology we follow for finding multivariate \AU relations using PCA.
\begin{enumerate}
    \item Select a number of explanatory variables $F_1, \ldots, F_n$ and a target feature $F$.
    \item Perform PCA on the feature set 
    \begin{equation}
    \mathbf{F} = \{F_1, \ldots, F_n, F\}\,.
    \end{equation}
    \item For each principal component $\mathbf{A_i}$, solve the equation 
    \begin{equation}
    \begin{split}
       \mathbf{A_i} \cdot \mathbf{F} &= \sum_{j=1}^n a_{i,j} F_j + a_{i,n+1} F \\
       \Rightarrow\qquad F &= -\left( \sum_{j=1}^n \frac{a_{i,j}}{a_{i,n+1}} F_j\right) \\
       \Rightarrow\qquad F &= \hat a_1 F_1 + \ldots + \hat a_n F_n
    \end{split}
    \label{eq:proj}
    \end{equation}
    with
    \begin{equation}
    \hat a_j = - \frac{a_{i,j}}{a_{i,n+1}},
    \end{equation}
    where we denote the right hand side as the new \emph{combined feature} $\hat F$
    \begin{equation}
    \hat F = \hat a_1 F_1 + \ldots + \hat a_n F_n \,.
    \label{eq:combined}
    \end{equation}
    \item Evaluate whether there exists a strong correlation between $F$ and a combined feature $\hat F$ using bivariate correlation analysis.
    \item If a strong correlation is found, choose a suitable model and fit it for the relation between $F$ and $\hat F$.
\end{enumerate}
In contrast to the bivariate case, this approach cannot be fully automated yet. A lot of guesswork is involved in identifying the principal components from which we can derive suitable combined features. The most straightforward approach for this task is to simply construct the combined feature for all principal components and then perform a bivariate correlation analysis of the target feature $F$ with each found combined feature.

Also, this method will not always yield universal relations: sometimes, there will be no principal component that will suitably explain the variance in the target feature $F$. This might happen in cases where a) $F$ simply does not present much variance across the whole data set, or b) there exist many co-linearities within the selected set of features $\mathbf{F}$. We discuss some cases where the method described above does not yield a \AU relation in Appendices~\ref{app:multivariate_special} and~\ref{app:multivariate_counter}.

\section{Multivariate Universal Relations}\label{sec:multivariate_relations}
We present the results of using PCA to find multivariate \AU relations for neutron stars as described in the previous section. A table summarizing all universal relations presented in this section can be found in the Conclusion~\ref{sec:conclusion}.

\subsection{Multivariate Universal Relations for Tidal Deformability}\label{sec:multivariate_tidal}
We here consider the case where we want to construct a \AU relation for the normalized tidal deformability $\bar\lambda$, using the features $M$, $R$ and $C$. To this end, we perform the principal component analysis on all 4 features using \textbf{Data set A} (cf. Section~\ref{sec:data}). The resulting principal components are given in Table~\ref{tab:multi_example} by means of the loading of each feature within the principal components. A visual representation of the combined feature obtained from each principal component is shown in Appendix~\ref{app:multivariate_details}.

As we can see in Table~\ref{tab:multi_example}, the target feature $\bar\lambda$ has the largest loading for Principal Component 2, with component 0 also showing a relatively large loading of $\bar\lambda$. Performing the bivariate correlation analysis of the combined features derived from each principal component with the target feature $\bar\lambda$ shows that the best correlation is actually given by Principal Component 0. However, through the visual inspection of the combined features, as depicted in Appendix~\ref{app:multivariate_details}, we can see that component 2 could also be leveraged for a universal relation, albeit with a larger error. For the remainder of the text we will focus on the relation with the lesser error, induced by component 0.

\begin{figure}
\centering
\includegraphics[width=0.9\columnwidth]{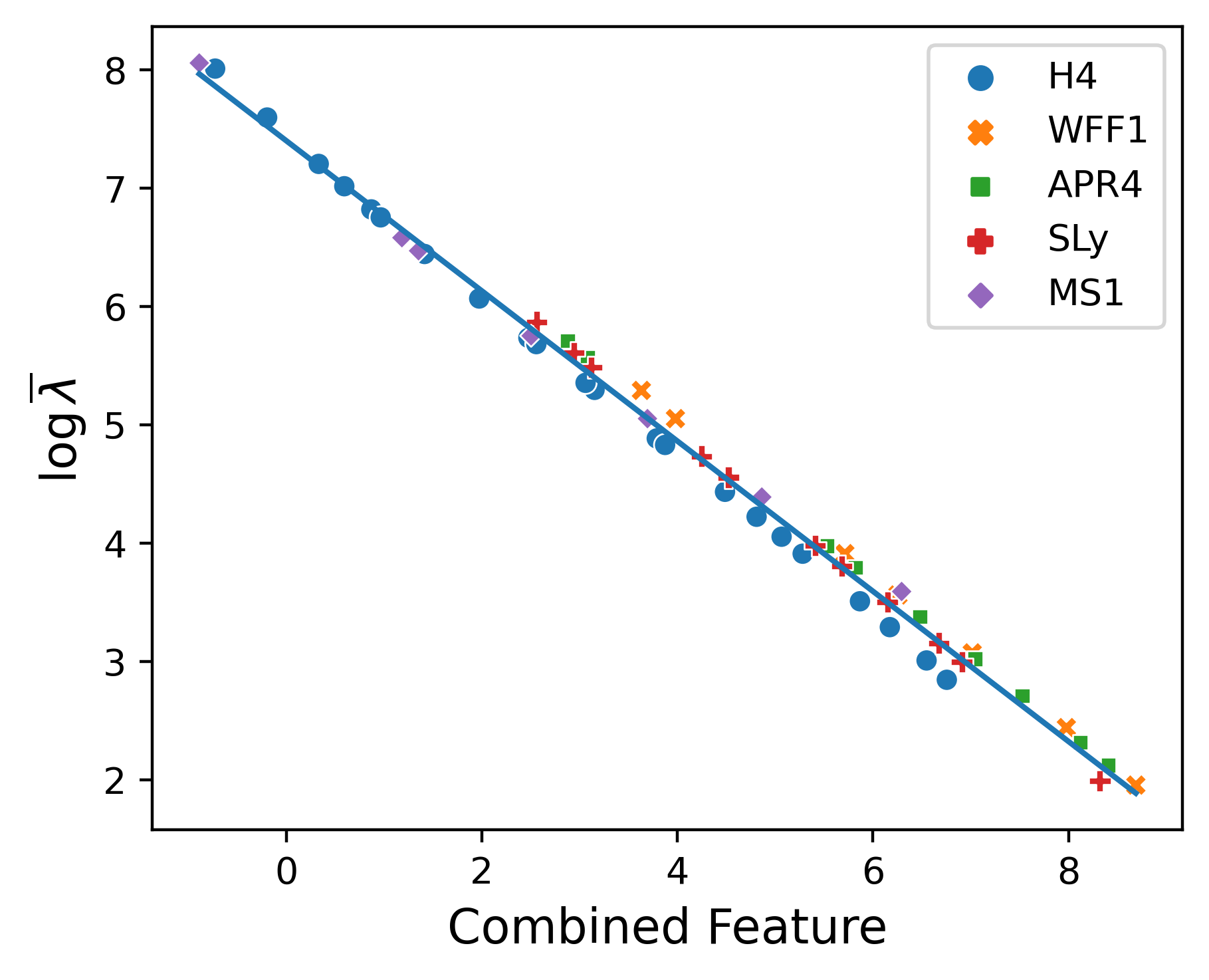}
\caption{Universal relation between logarithm of the normalized tidal deformability $\log \bar\lambda$ and the combined feature given in Equation~\eqref{eq:combined_tidal}, derived from Principal Component 0 (cf. Table~\ref{tab:multi_example}), using \textbf{Data set A}~\cite{2021PhRvD.104b3005M,kruger2019fast}.}
\label{fig:multi_fit}
\end{figure}

Since the automated approach finds an exponential relation between $\bar \lambda$ and the combined feature, and we again fit for $\log \bar \lambda$ to obtain a more accurate relation. 
Through our manual fit, we obtain the following \AU relation for the normalized tidal deformability
\begin{equation}
\log \bar \lambda = \num{-0.63455955} \hat F +\num{7.39926917}
\label{eq:tidal_relation}
\end{equation}
with 
\begin{equation}
\hat F = \num{3.390534206245} \frac{M}{M_\odot} - \num{5.2414369999999995} \frac{R}{10 km} + \num{4.768246}\frac{C}{0.2}.
\label{eq:combined_tidal}
\end{equation}
This relation is presented in Figure~\ref{fig:multi_fit} and achieves an average relative error of $\num{0.023403790657464087}$. 
Compared to the bivariate relation between the tidal deformability and compactness we presented in Figure~\ref{fig:bivariate_2}, we essentially introduce a linear order correction involving the radius and the mass. While the overall relative error is approximately the same as for the bivariate relation, the multivariate relation remains entirely linear in all independent variables, reducing its sensitivity to potential estimation errors for these quantities.

\begin{table}
\def\arraystretch{1.5}%
\centering
\begin{tabular}{c | r | r | r | r | r}
Component & $M$ & $R$ & $C$ & $\bar \lambda$ \\\hline
0 & \num{-0.48793098} & \num{0.41864877} & \num{-0.57054369}  & \num{0.51101514} \\
1 & \num{ 0.59621943}  & \num{0.7932779}  & \num{-0.03405559} & \num{-0.11862875} \\
2 & \num{ 0.32170168} & \num{-0.09744886} & \num{ 0.41194891}  & \num{0.84694147} \\
3 & \num{ 0.55041238} & \num{-0.43121584} & \num{-0.70966063}  & \num{0.08649222}
\end{tabular}
\caption{Loadings of features in each principal component obtained from performing PCA on the feature set $\mathbf{F} = \{M, R, C, \bar\lambda\}$ on \textbf{Data set A}.}
\label{tab:multi_example}
\end{table}

\medskip
\noindent
\textbf{Relation with \textbf{Data set B}.}
We also perform the same analysis using the data by Kuan et al.~\cite{Kuan:2021jmk}. The principal components obtained from the PCA are listed in Table~\ref{tab:multi_example_Kuan}. The principal components show a similar behavior to the previous examples using \textbf{Data set A}, however we can observe some slight differences caused by the different equations of state used in the data set. 

\begin{figure}
\centering
\includegraphics[width=0.9\columnwidth]{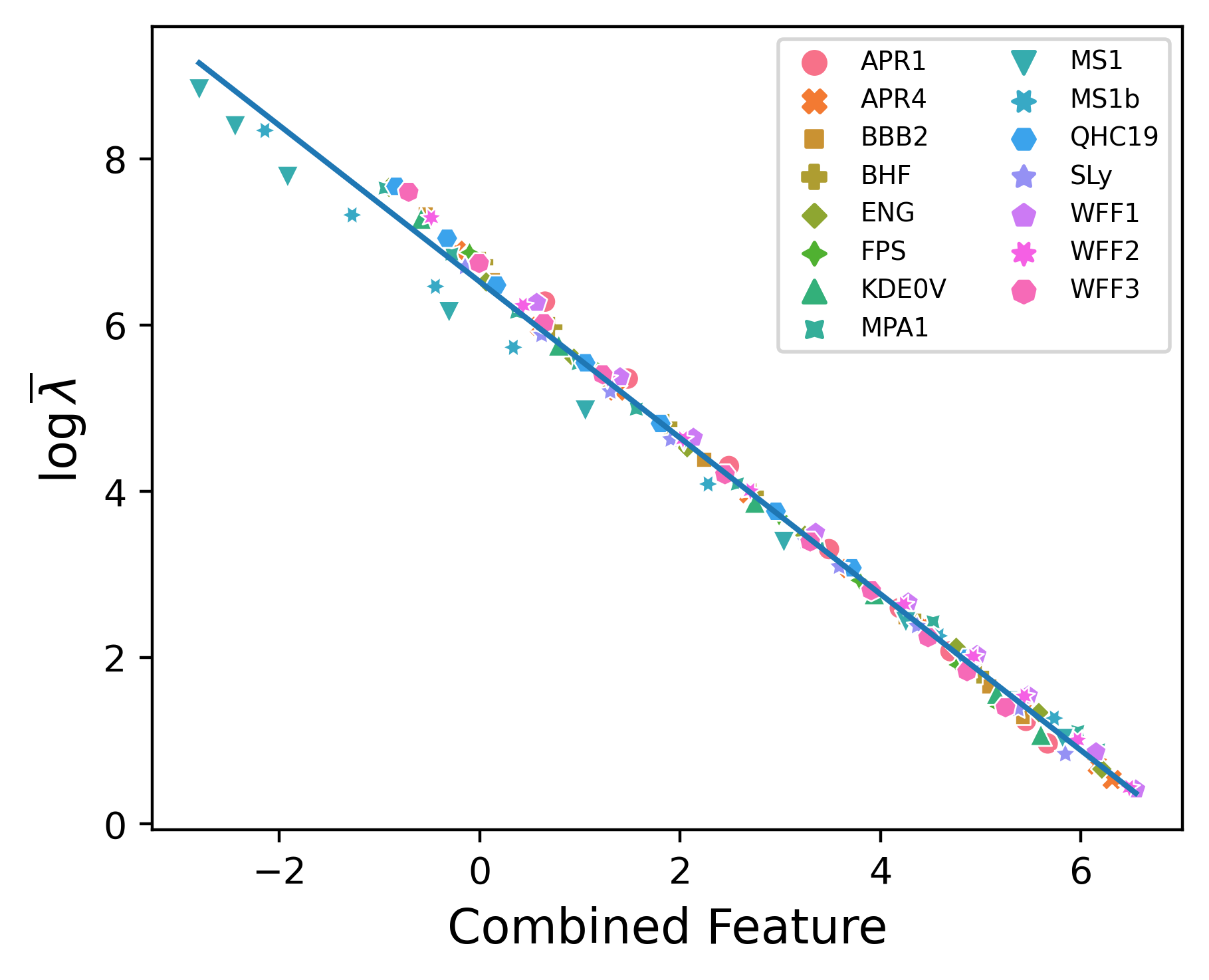}
\caption{Universal relation between the logarithm of the normalized tidal deformability $\log \bar\lambda$ and the combined feature given in Equation~\eqref{eq:tidal_relation_Kuan}, derived from Principal Component 0 (cf. Table~\ref{tab:multi_example_Kuan}), using \textbf{Data set B}~\cite{Kuan:2021jmk}.}
\label{fig:multi_fit_Kuan}
\end{figure}

As before, after performing the bivariate correlation analysis on the combined features derived from each principal component, we find that the combined feature derived from Principal Component 0 shows the best universality. Leveraging this component, we obtain the \AU relation 
\begin{equation}
\log \bar \lambda = \num{-0.93929701} \hat F + \num{6.52146358}
\label{eq:tidal_relation_Kuan2}
\end{equation}
this time with the combined feature
\begin{equation}
\hat F = \num{2.24908701354} \frac{M}{M_\odot} - \num{4.315921} \frac{R}{10 km} + \num{3.5326000000000004}\frac{C}{0.2}\,.
\label{eq:tidal_relation_Kuan}
\end{equation}
The resulting best fit is presented in Figure~\ref{fig:multi_fit_Kuan}. It achieves an average relative error of $\num{0.042847553820507985}$, which is slightly higher than what we achieved for \textbf{Data set A}. We suspect this is caused by some of the outlying neutron star models that are introduced by the larger configuration space considered in \textbf{Data set B}. 

However, the fact remains that our approach for the multivariate correlation analysis yields the same form for the universal relation independent of which data set is used. This is indicative of this approach further generalizing well for different data sets, and that the results presented here are not dependent on the underlying data used for the analysis.

\begin{table}
\def\arraystretch{1.5}%
\centering
\begin{tabular}{c | r | r | r | r | r}
Component & M & R & $C$ & $\bar \lambda$ \\\hline
0 & \num{-0.5142147 } & \num{0.34755964} & \num{-0.59262472} & \num{0.51340187} \\
1 & \num{0.55470435} & \num{0.79888905} & \num{0.14483504} & \num{0.18194005} \\
2 & \num{0.12718581} & \num{-0.35179198}  & \num{0.40576465}  & \num{0.83391919} \\
3 & \num{-0.6416464}  & \num{ 0.3423755}   & \num{0.68056873} & \num{-0.08885448}
\end{tabular}
\caption{Loadings of features in each principal component obtained from performing PCA on the feature set $\mathbf{F} = \{M, R, C, \bar\lambda\}$ on \textbf{Data set B}.}
\label{tab:multi_example_Kuan}
\end{table}

\subsection{Multivariate Astroseismological Relations}\label{sec:multivariate_oscillation}
Andersson and Kokkotas~\cite{PhysRevLett.77.4134,Andersson98} previously proposed a \AU relation linking the average density $\tilde\rho$ to the $f$-mode frequency of a neutron star. We here attempt to apply the same method as above to potentially find corrections to their original astroseismological relation that improve its universality. To this end, we perform the principal component analysis on the features $\omega_f$, $M$, $C$ and $\tilde\rho$, aiming at finding corrections in terms of $M$ and $C$ for the \AU relation.

\begin{figure}
\centering
\includegraphics[width=0.9\columnwidth]{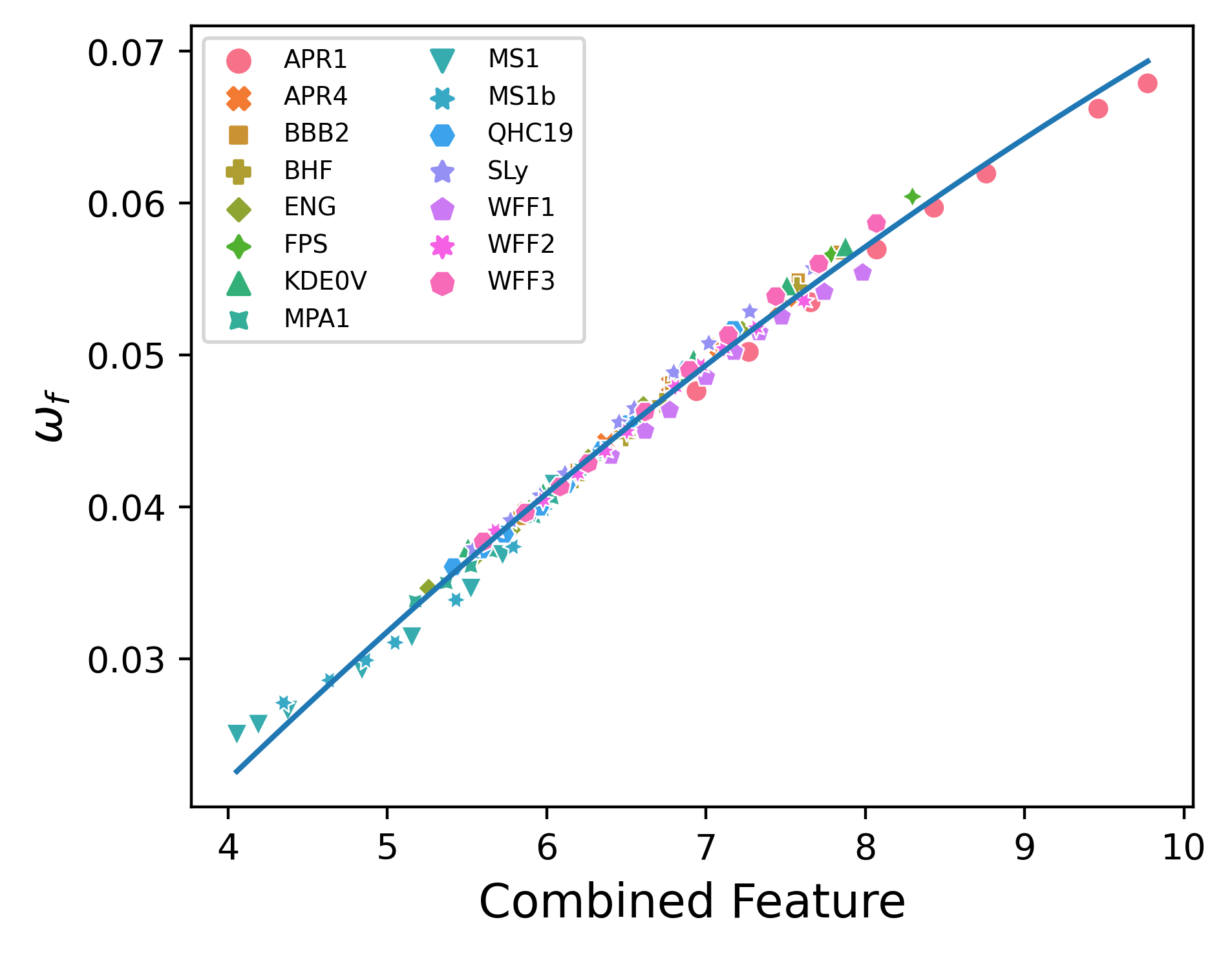}
\caption{Universal relation for the $f$-mode frequency $\omega_f$ using the combined feature of $M$, $\tilde\rho$ and $C$ given in Equation~\eqref{eq:combined_fmode}, obtained from the PCA on \textbf{Data set B}~\cite{Kuan:2021jmk}.}
\label{fig:multi_fit_fmode}
\end{figure}

The best relation is found for the combined feature derived from the fourth principal component found through PCA performed in the feature set $\mathbf{F} = \{M, C, \tilde\rho, \omega_f\}$. . The best fit for the relation between $\omega_f$ and this combined feature is shown in Figure~\ref{fig:multi_fit_fmode}. The best fit shows a quadratic \AU relation for the $f$-mode frequency of the form
\begin{equation}
\omega_f = -0.00033 \hat F^2 + \num{0.012682} \hat F - \num{0.02348942}
\label{eq:relation_fmode}
\end{equation}
with
\begin{equation}
\hat F = \num{2.9800476501500004} \frac{M}{M_\odot} + \num{10.230908}\frac{\tilde\rho}{0.04} - \num{8.39804} \frac{C}{0.2}\,.
\label{eq:combined_fmode}
\end{equation}
This relation achieves an average relative error of $\num{0.015469332922977337}$. When compared to the old relation shown in Figure~\ref{fig:bivariate3_0}, we can clearly observe an improved universality, which is also reflected in the average relative error that is reduced by half. We thus achieve a significant improvement over the existing relation by using our multivariate approach.

\subsection{Improved Astroseismological Relations for the f-mode Frequency}\label{sec:multivariate_example3}
We next consider another variation on the astroseismological relation we inspected above. This time, instead of introducing mass and compactness as independent variables, we instead only introduce the product $C \tilde\rho$ of compactness and average density as a new independent variable.
Our goal now is therefore to find a \AU relation for $\omega_f$ using the average density $\tilde\rho$ and $C \tilde\rho$. 

\begin{figure}
\centering
\includegraphics[width=0.9\columnwidth]{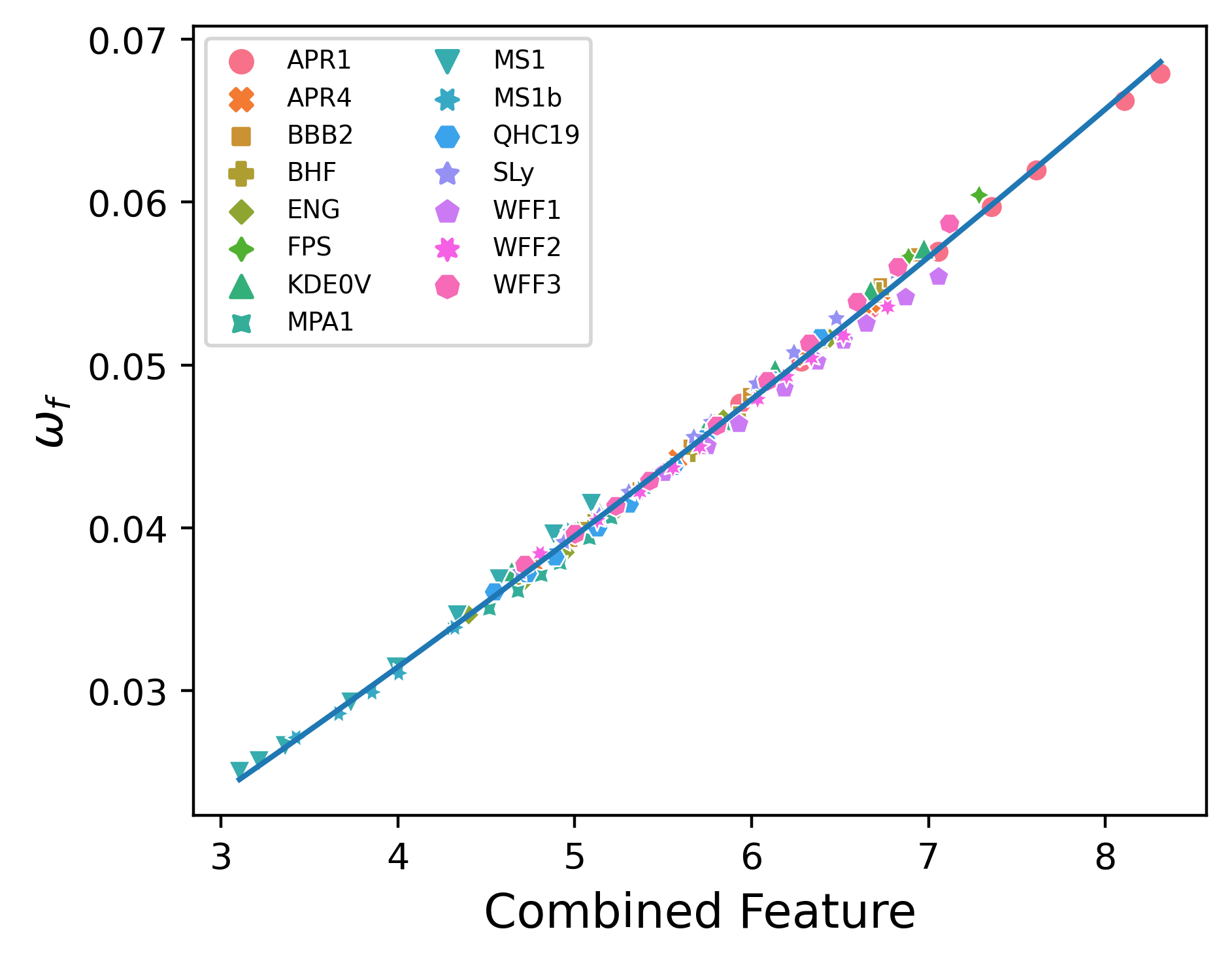}
\caption{Universal relation for the $f$-mode frequency $\omega_f$ using the combined feature of $\tilde\rho$ and $C \tilde\rho$ given in Equation~\eqref{eq:combined_feature_imp_zmode}, obtained from the PCA on \textbf{Data set B}~\cite{Kuan:2021jmk}.}
\label{fig:multi_fit_zmode}
\end{figure}

In this case, the best relation is found for the combined feature derived from the third principal component found through the PCA performed in the feature set $\mathbf{F} = \{\tilde\rho, C \tilde\rho, \omega_f\}$. 
The best fit for the relation between $\omega_f$ and this combined feature is shown in Figure~\ref{fig:multi_fit_zmode}. The best fit shows a quadratic \AU relation for the $f$-mode frequency of the form
\begin{equation}
\omega_f = 0.0002 \hat F^2 + \num{0.00647675} \hat F + \num{0.00276012}
\label{eq:multivariate_relation_redshift}
\end{equation}
with
\begin{equation}
\hat F = \num{6.9109313199999995}\frac{\tilde\rho}{0.04} - \num{1.71649574} \frac{C \tilde\rho}{0.01}\,.
\label{eq:combined_feature_imp_zmode}
\end{equation}

When compared to the relation shown in the section above (cf. Figure~\ref{fig:multi_fit_fmode}) we observe an improved universality: the previous relation has an average relative error of $\num{0.015469332922977337}$, whereas the relation with the new combined feature achieves an error of $\num{0.009738598872994107}$. 

Considering that the original relation put forward by Andersson and Kokkotas~\cite{PhysRevLett.77.4134,Andersson98} was inspired by Newtonian gravity, the additional factor in $C \tilde\rho$ could be considered a first order correction to account for general relativity, since
\begin{equation}
\hat F = \num{172.773283} \tilde\rho - \num{171.649574}C \tilde\rho \approx 172 \tilde\rho \left(1-C\right)\,.
\end{equation}
Essentially, this new relation is a stepping-stone between the relation by Andersson and Kokkotas~\cite{PhysRevLett.77.4134,Andersson98}, and other general relativistic \AU relations, such as the one between the $f$-mode frequency $\omega_f$ and the compactness $C$ put forward by Tsui and Leung~\cite{PhysRevLett.95.151101}.

\subsection{Discussion of Results}
As we have demonstrated above, we can utilize the principal components obtained from PCA to construct multivariate \AU relations for neutron stars. Since the relations we construct are, for now, first-order relations, this approach is also suited for finding first-order corrections to existing \AU relations, allowing an improvement of the accuracy of the \AU relations.

Despite these positive results, our approach here has only been descriptive: while we provide a methodology that can yield multivariate \AU relations, the formal reasons for why this approach works is still not fully clear. Gaining further understanding of the mathematical underpinnings of this approach can allow us to further improve its output, but also better understand its limits. 

For instance, our findings do not agree with the observations made in~\cite{2021A&A...654A.162S}: they claim that the best relations would follow from the last principal component obtained through the principal component analysis. In our findings, however, the best universal relations can appear from any of the principal components. It is therefore our belief that further analysis of the PCA method and the structure of its principal components is necessary to obtain a more rigorous understanding of this approach. Until then, the PCA approach should only be used to generate potential candidate relations that have to be further analyzed for their accuracy.

We have also seen that it is not always the principal component with the largest loading for our target feature that will induce the best universal relation: in the example discussed in Section~\ref{sec:multivariate_tidal}, we had two principal components with relatively large loadings for the target feature, but ultimately the combined feature derived from the principal component with the second largest loading offered the relation with the smallest error. A test of all possible combined features using the bivariate analysis therefore remains necessary until we potentially devise alternative criteria for deciding which principal component should be used to induce a universal relation.

Finally, in Appendices~\ref{app:multivariate_special} and~\ref{app:multivariate_counter}, we show some cases where our approach will not yield any \AU relations. Sometimes this is caused by the data used, as, ultimately, not all feature combinations will be amenable to \AU relations. Furthermore, specific properties of the used data, such as the existence of strong collinearities with the target feature, can also hinder our approach from producing \AU relations. We currently can only provide superficial reasons for why our approach does not perform well in such situations, and we hope to obtain a more rigorous understanding through future work. 

\begin{table*}
\centering
\def\arraystretch{2}
\begin{tabularx}{\linewidth}{l | c | X | c | c | c | c}
Type & Features & Form & Avg. rel. Error & Equation & Figure & Reference \\\hline
\multirow{9}{*}{\rotatebox[origin=c]{90}{\Large{Bivariate}}}
& $\bar \lambda$, $\bar I$ & $\bar I = \num{0.01860855} \log \bar\lambda^2 - \num{0.07584032}\log \bar\lambda + \num{0.33375708}$ & $\num{0.020266944213251526}$ & \eqref{eq:bivariate_0} & \ref{fig:bivariate_0} & \cite{Yagi365} \\
& $\bar \lambda$, $\eta$    & $\log \bar \lambda = \num{-0.09253471}\eta^2 -\num{5.42472023} \eta +\num{13.60418765}$            & $\num{0.007502394271055736}$ & \eqref{eq:bivariate_1} & \ref{fig:bivariate_1} & \cite{2021PhRvD.104b3005M}\\
& $\bar \lambda$, $C$ & $\log \bar \lambda =  \num{46.1226271} C^2 -\num{53.04505671} C + \num{13.63341181}$ & $\num{0.019736506644986718}$ & \eqref{eq:bivariate_2} & \ref{fig:bivariate_2} & \cite{Jiang_2020} \\
& $\bar M \omega_f$, $C$ & $\bar M \omega_f = \num{0.04212737} \log C^2+ \num{0.22183278} \log C + \num{0.3145776}$ & $\num{0.01148544725114905}$ & \eqref{eq:bivariate2_0} & \ref{fig:bivariate2_0} & \cite{PhysRevLett.95.151101} \\
& $\bar M \omega_f$, $\bar I$ & $\bar M \omega_f = \num{0.02130106} \log {\bar I}^2 - \num{0.01956669} \log \bar I + \num{0.03207015}$ & $\num{0.006879103949206297}$ & \eqref{eq:bivariate2_1} & \ref{fig:bivariate2_1} & /\\
& $\bar M \omega_f$, $\bar \lambda$ & $\bar M \omega_f = 0.0003 \log {\bar \lambda}^2 - \num{0.01490679} \log \bar \lambda + \num{0.12746998}$ & $\num{0.01425718984920659}$ & \eqref{eq:bivariate2_2} & \ref{fig:bivariate2_2} & \cite{chan2014multipolar}\\
& $\bar M \omega_f$, $\eta$ & $\bar M \omega_f = \num{0.0149526} \eta^2 + \num{0.02475} \eta - \num{0.00885211}$ & $\num{0.006629639167064198}$ &\eqref{eq:bivariate2_3} & \ref{fig:bivariate2_3} & \cite{lau2010inferring,PhysRevD.91.044034,kruger2019fast} \\
& $\omega_f$, $\tilde\rho$ & $\omega_f = \num{-2.19908806} \tilde\rho^2 + \num{0.98518922} \tilde\rho + \num{0.007272}$ & $\num{0.03499534511917183}$ & \eqref{eq:bivariate3_0} & \ref{fig:bivariate3_0} & \cite{PhysRevLett.77.4134,Andersson98,PhysRevD.70.124015}\\
& $\bar M \omega_{g_1}$, $R \omega_f$ & $\log \bar M\omega_{g_1} = \num{16.05167444} \left(R \omega_f\right)^2 -\num{5.32343308} R \omega_f+ \num{5.58908468}$ & $\num{0.004346355091302037}$ & \eqref{eq:bivariate2_5} & \ref{fig:bivariate2_5} & \cite{Kuan:2021jmk} \\\hline
\multirow{3}{*}{\rotatebox[origin=c]{90}{\Large{Multivariate\quad}}}
& $\bar\lambda$, $M$, $R$, $C$ & $\log \bar \lambda = \num{-0.63455955} \hat F +\num{7.39926917}$ \newline $\hat F = \num{3.390534206245} \frac{M}{M_\odot} - \num{5.2414369999999995} \frac{R}{10 km} + \num{4.768246}\frac{C}{0.2}$ & $\num{0.023403790657464087}$ & \eqref{eq:tidal_relation} & \ref{fig:multi_fit} & / \\
& $\omega_{f}$, $M$, $\tilde\rho$, $C$ & $\omega_f = -0.00033 \hat F^2 + \num{0.012682} \hat F - \num{0.02348942}$ \newline $\hat F = \num{2.9800476501500004} \frac{M}{M_\odot} + \num{10.230908}\frac{\tilde\rho}{0.04} - \num{8.39804} \frac{C}{0.2}$ & $\num{0.015469332922977337}$ & \eqref{eq:relation_fmode} & \ref{fig:multi_fit_fmode} & / \\
& $\omega_{f}$, $\tilde\rho$, $C\tilde\rho$ & $\omega_f = 0.0002 \hat F^2 + \num{0.00647675} \hat F + \num{0.00276012}$\newline $\hat F = \num{6.9109313199999995}\frac{\tilde\rho}{0.04} - \num{1.71649574} \frac{C \tilde\rho}{0.01}$ & $\num{0.009738598872994107}$ & \eqref{eq:multivariate_relation_redshift} & \ref{fig:multi_fit_zmode} & / \\\hline
\end{tabularx}
\caption{List of all universal relations presented in this work.}
\label{tab:relations}
\end{table*}

\section{Conclusion \& Future Directions}\label{sec:conclusion}
In this work, we discussed the potential of approaching the task of constructing \emph{universal} relations for neutron stars from a statistical data analysis point of view. Instead of relying on physical intuition, our goal was to approach neutron star data using statistical methods only and thus enable a more automated approach to finding \AU relations. 

In a first step, we investigated the suitability of four different correlations measures for identifying pairs of features amenable to bivariate \AU relations. We found that the usual Pearson correlation measure will have difficulties with non-linear relations between features, which has also been observed in the past in the statistical data analysis literature for more general use cases~\cite{clark2013comparison}. Using generalized correlation measures that were explicitly constructed to detect non-linear correlations proved more useful: overall, Mutual Information and Maximal Information both performed best in finding universally related features, and while Distance correlation did not perform as well as the aforementioned ones, it still outperformed Pearson correlation for our use case. 

In a second step, we also approached the problem of constructing multivariate \AU relations. Inspired by an idea presented in~\cite{2021A&A...654A.162S}, we used the principal components found through PCA to construct a new combined feature that we then related to a initially selected target feature. While this approach is not yet fully automated and requires manual considerations in some steps, our results show that this approach can yield highly accurate, multivariate \AU relations. Our approach works particularly well when we try to find first-order corrections to previously known bivariate relations. For instance, we were able to construct an entirely novel \AU relation that allows us to relate the $f$-mode frequency to the average density and compactness of the neutron stars, significantly improving the error of the relation compared to existing bivariate relations.

In Table~\ref{tab:relations} we give an overview of all universal relations presented in this paper. For each relation, we indicate which features are connected through these relations, their form, and the average relative error achieved through our best fits. We also give references to all corresponding equations and figures in this paper. Finally, if a relation was already presented previously in a different work, we also give a reference to that work.

In a time where theoretical model data for various (astro-)physical objects becomes more widely available, finding useful data analysis tools for the specific use-cases that we are interested in will be an important direction of work that will later enable more comprehensive data exploration. The methods discussed in this paper present a first step into this direction.

For future work, a straightforward extension is the application of the presented methods to even more and different neutron star data. While we have only considered non-rotating neutron stars in this paper, the presented methods should easily apply to other configurations including rotation or magnetic fields. Furthermore, gaining deeper understanding on why and under which constraints the PCA approach will work well can allow us to, in the future, reduce the amount of manual intervention that is still required right now.

\bibliographystyle{apsrev4-2}

\onecolumngrid
\clearpage
\twocolumngrid
\appendix
\section{Combined Features from Multivariate Correlation Analysis}\label{app:multivariate_details}
\begin{figure*}
\centering
\subfloat[][Principal Component 0]{\includegraphics[width=0.5\textwidth]{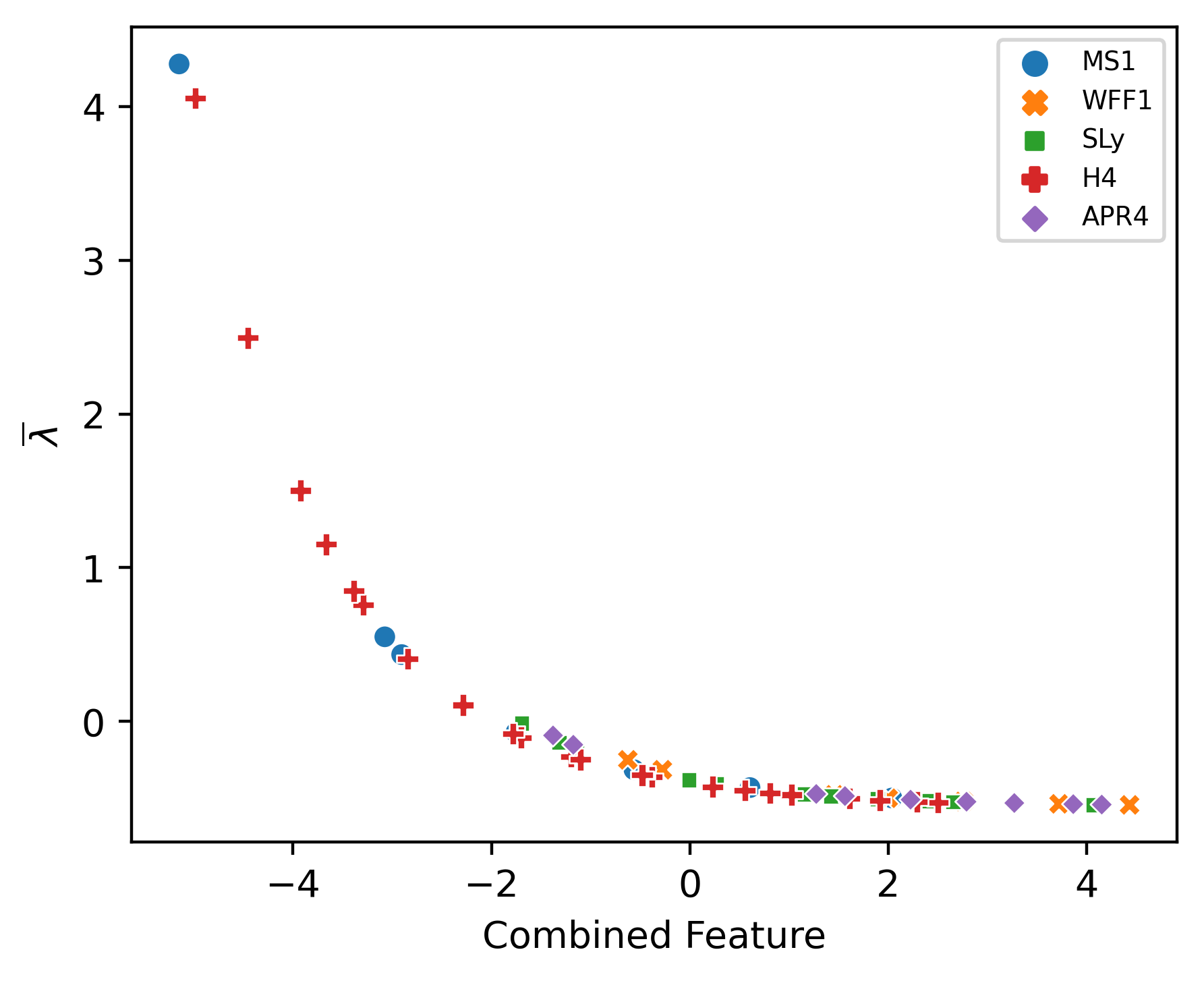}\label{fig:multi_0}}
\subfloat[][Principal Component 1]{\includegraphics[width=0.5\textwidth]{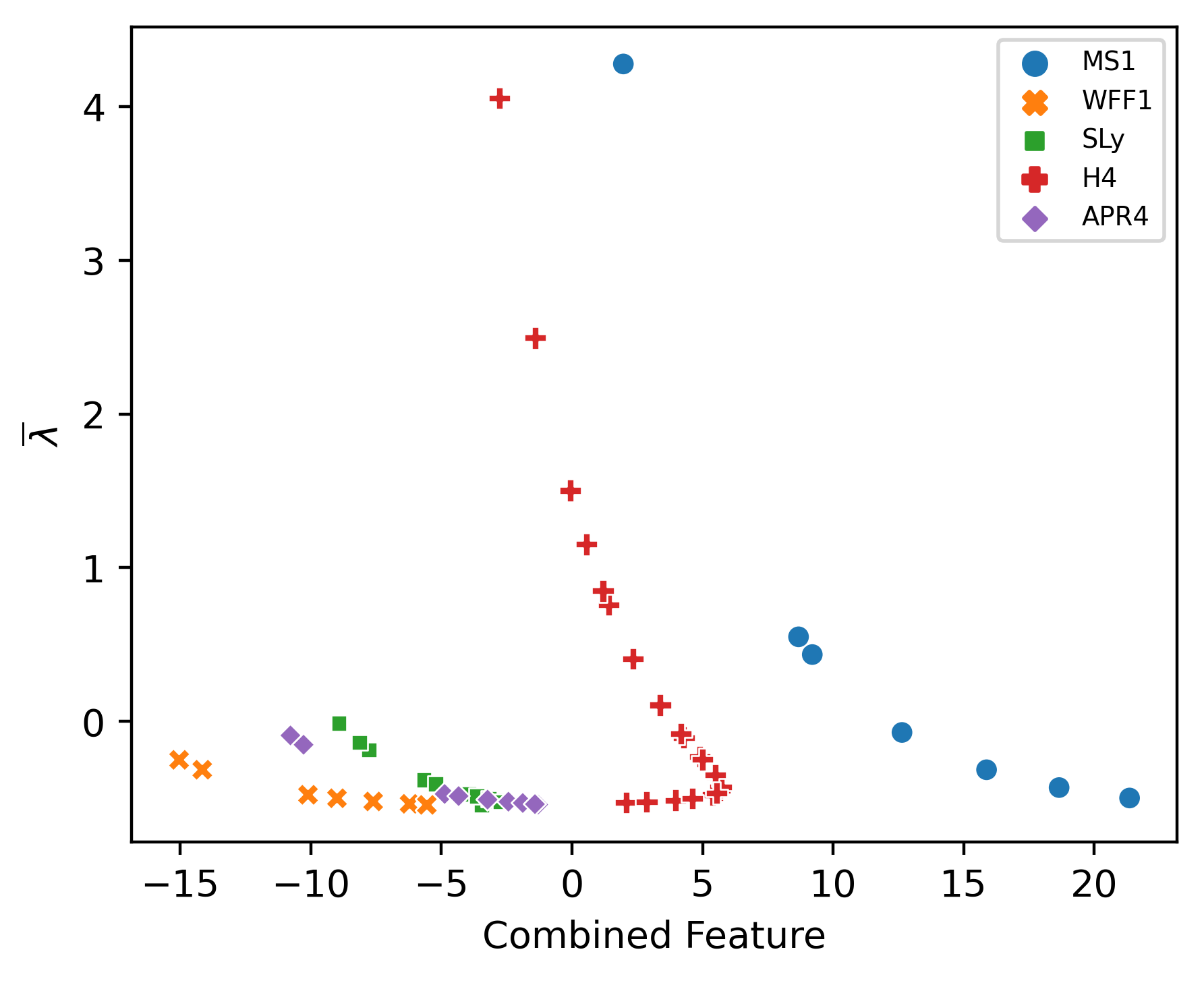}\label{fig:multi_1}}

\subfloat[][Principal Component 2]{\includegraphics[width=0.5\textwidth]{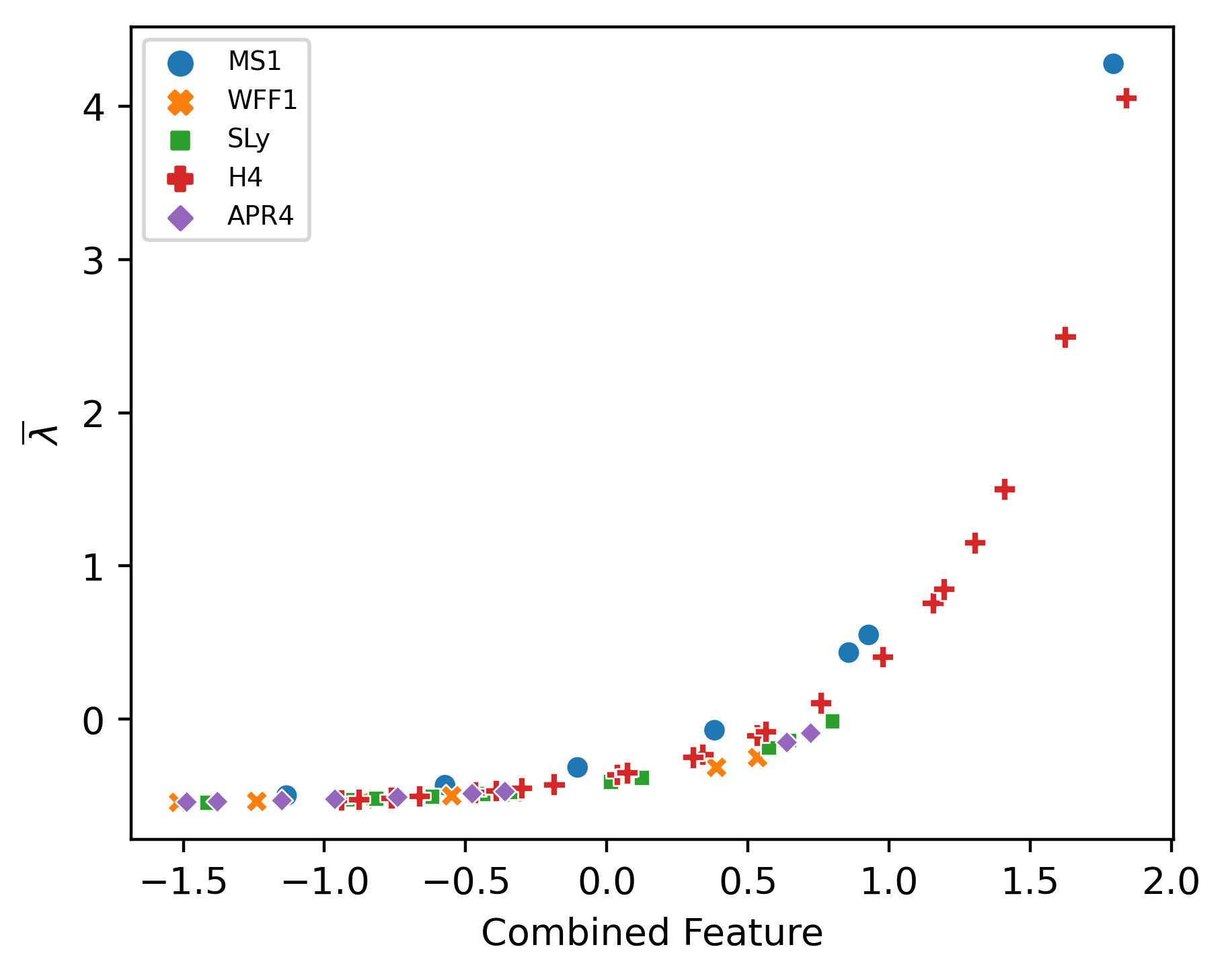}\label{fig:multi_2}}
\subfloat[][Principal Component 3]{\includegraphics[width=0.5\textwidth]{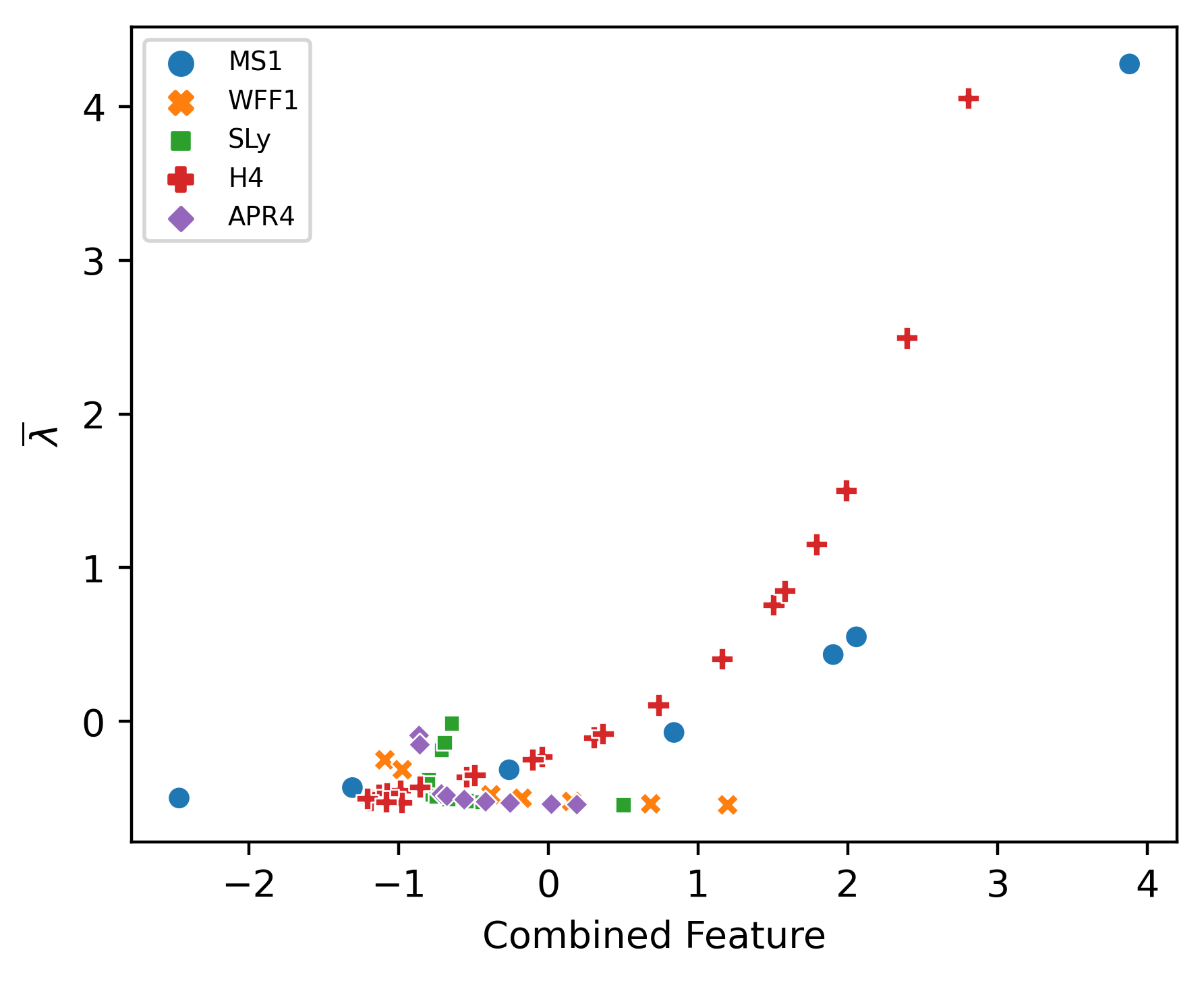}\label{fig:multi_3}}
\caption{The combined feature derived from each principal component solved for the target feature $\lambda$ after performing PCA on the feature set $\mathbf{F}=\{M, R, C, \lambda\}$. The corresponding loadings are given in Table~\ref{tab:multi_example}.}
\label{fig:multi_example}
\end{figure*}

We here show in Figure~\ref{fig:multi_example} a visual representation of correlating the combined features we obtained in Section~\ref{sec:multivariate_tidal} with the target feature $\bar\lambda$. From this figures we can clearly see the strong correlation of the combined feature obtained from both Principal Component 0 and 2 with $\bar \lambda$. While the purely visual inspection already points towards Principal Component 0 allowing for the smaller error, a precise analysis using the bivariate correlation method was ultimately necessary to decide which component induces the universal relation with the least error. However, a similar visual analysis can and should be performed to assist in any attempt to construct \AU relations using multivariate data analysis.

\begin{table}[t]
\def\arraystretch{1.5}%
\centering
\begin{tabular}{c | r | r | r}
Component & $\eta$ & $\bar M \omega$& $C$ \\\hline
0 & \num{-0.57720817} & \num{0.57768953}& \num{-0.57715296}  \\
 1 & \num{-0.68722846} & \num{-0.03811131} & \num{0.72544095}\\
 2 & \num{-0.44107569} & \num{0.81536638}& \num{-0.37500653} \\
\end{tabular}
\caption{Loadings of features in each principal component shown in Figure~\ref{fig:multi_example3}}
\label{tab:multi_example3}
\end{table}

\section{The Special Case with strong Collinearity}\label{app:multivariate_special}
Unfortunately, the approach using multivariate statistical analysis we described in this work (cf. Section~\ref{sec:multivariate}) does not always produce conclusive results: in cases where there exist strong correlations between features, the conditions we formulated in Section~\ref{sec:multivariate_methodlogy} will not necessarily or sufficiently lead to the construction of \AU relations.

\begin{figure*}
\centering
\subfloat[][Principal Component 0]{\includegraphics[width=0.5\textwidth]{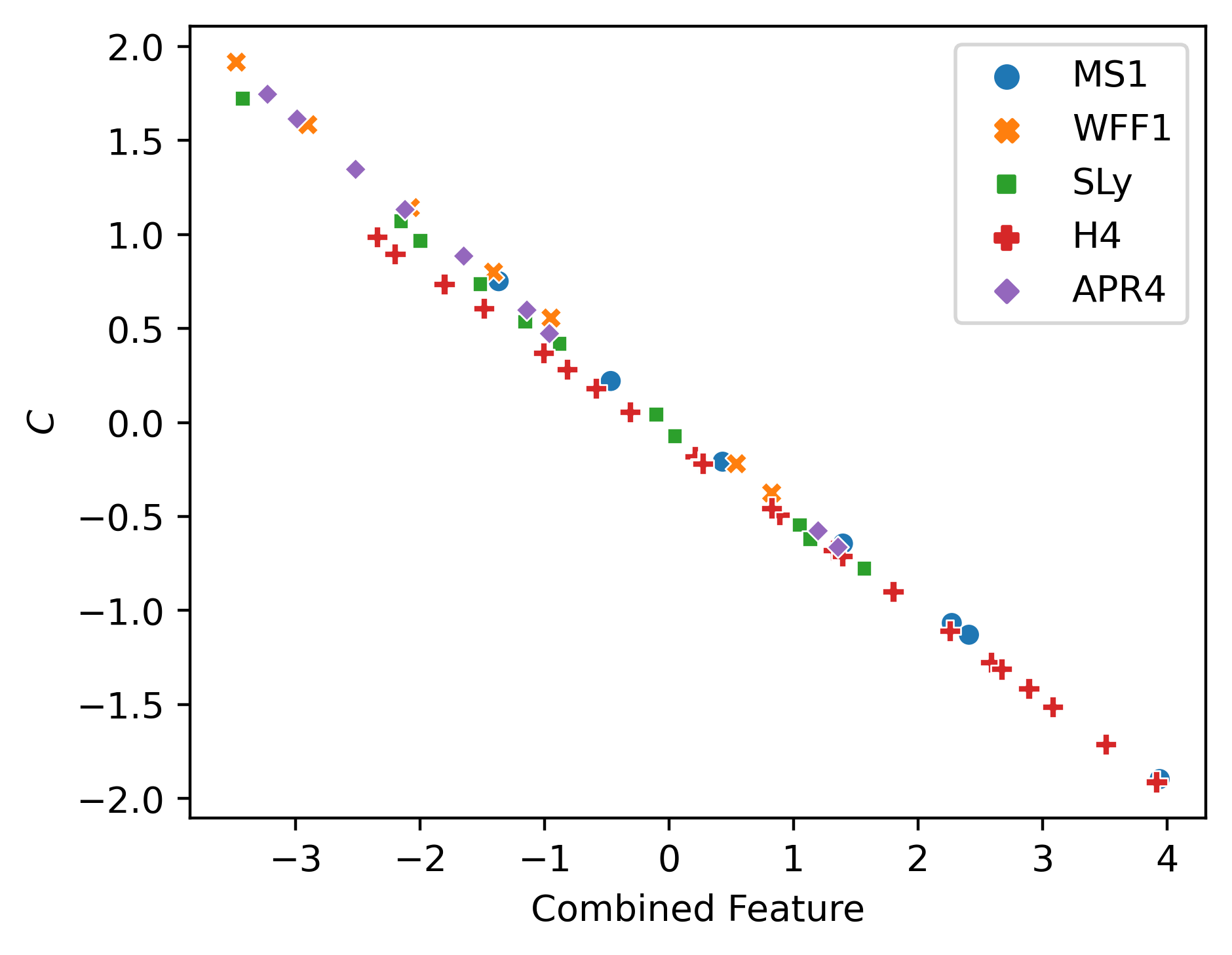}\label{fig:multi3_0}}
\subfloat[][Principal Component 1]{\includegraphics[width=0.5\textwidth]{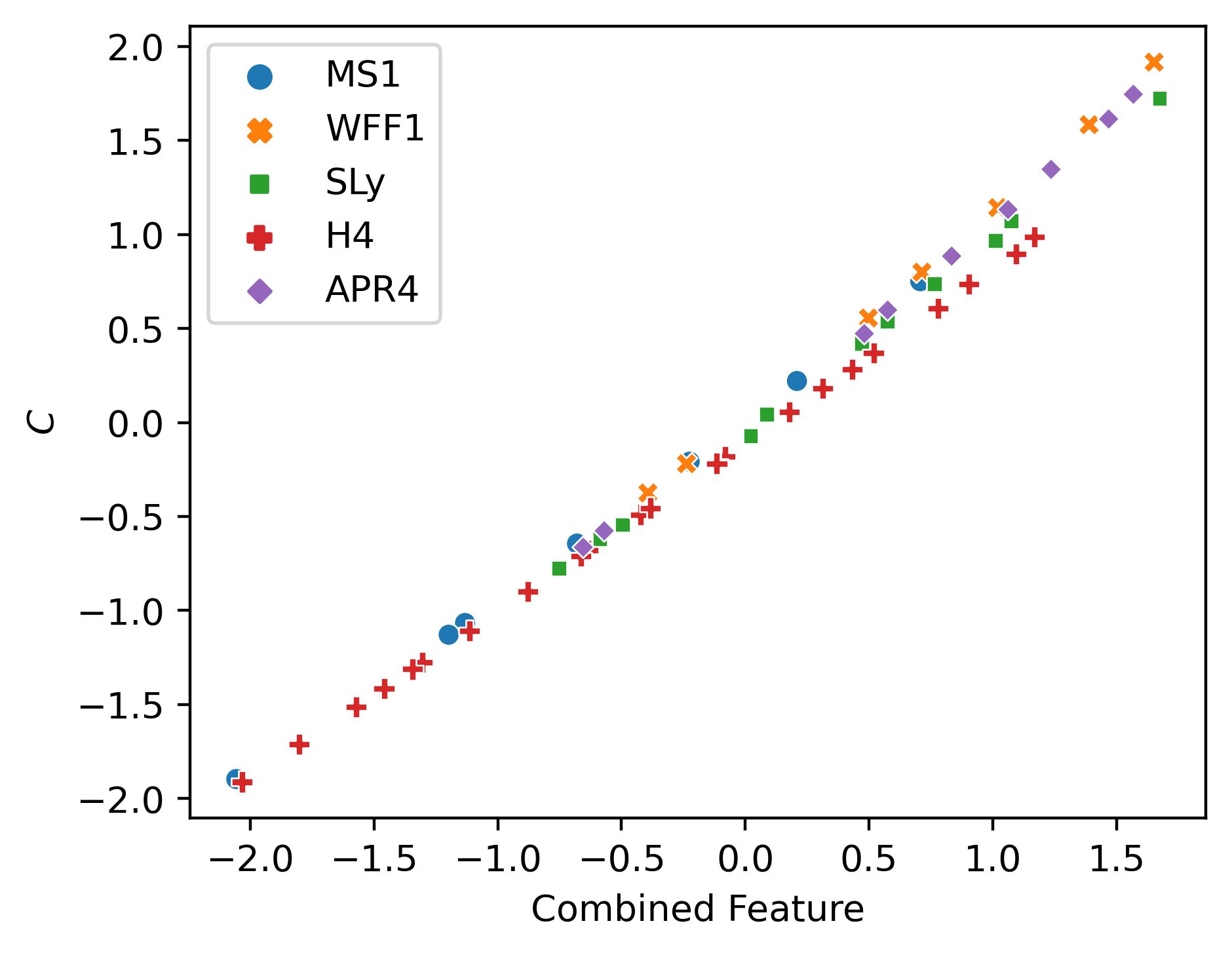}\label{fig:multi3_1}}

\subfloat[][Principal Component 2]{\includegraphics[width=0.5\textwidth]{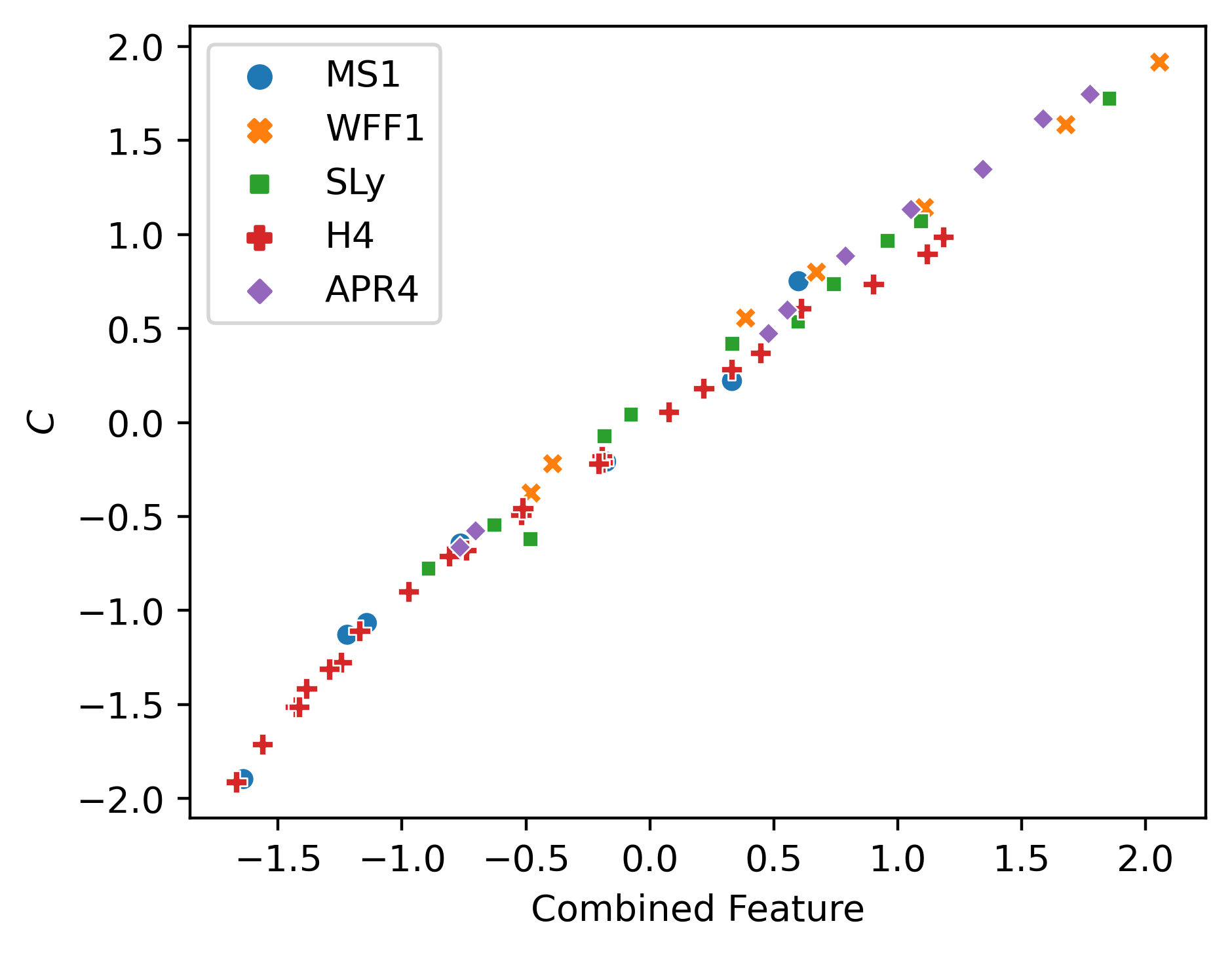}\label{fig:multi3_2}}
\caption{The combined feature derived from each principal component solved for the target feature $C$ after performing PCA on the feature set $\mathbf{F}=\{\bar M\omega_f, \eta, C\}$. The corresponding loadings are given in Table~\ref{tab:multi_example3}.}
\label{fig:multi_example3}
\end{figure*}

For instance, let's consider the case where we want to predict the compactness $C$ given the features $\bar M\omega$ and $\eta$. The principal component analysis leads to the loadings given in Table~\ref{tab:multi_example3}, and the associated combined features shown in Figure~\ref{fig:multi_example3}. As we can see, each corresponding combined feature is strongly correlated to $C$, however inspection of the loading does not necessarily yield any specific principal component for which $C$ has a significantly larger contribution. As such, not finding principal component with a proportionally large loading for our target feature does necessarily imply that no potential universal relation exists.

\section{Counterexample for Multivariate Correlation Analysis}\label{app:multivariate_counter}
\begin{figure*}
\centering
\subfloat[][Principal Component 0]{\includegraphics[width=0.5\textwidth]{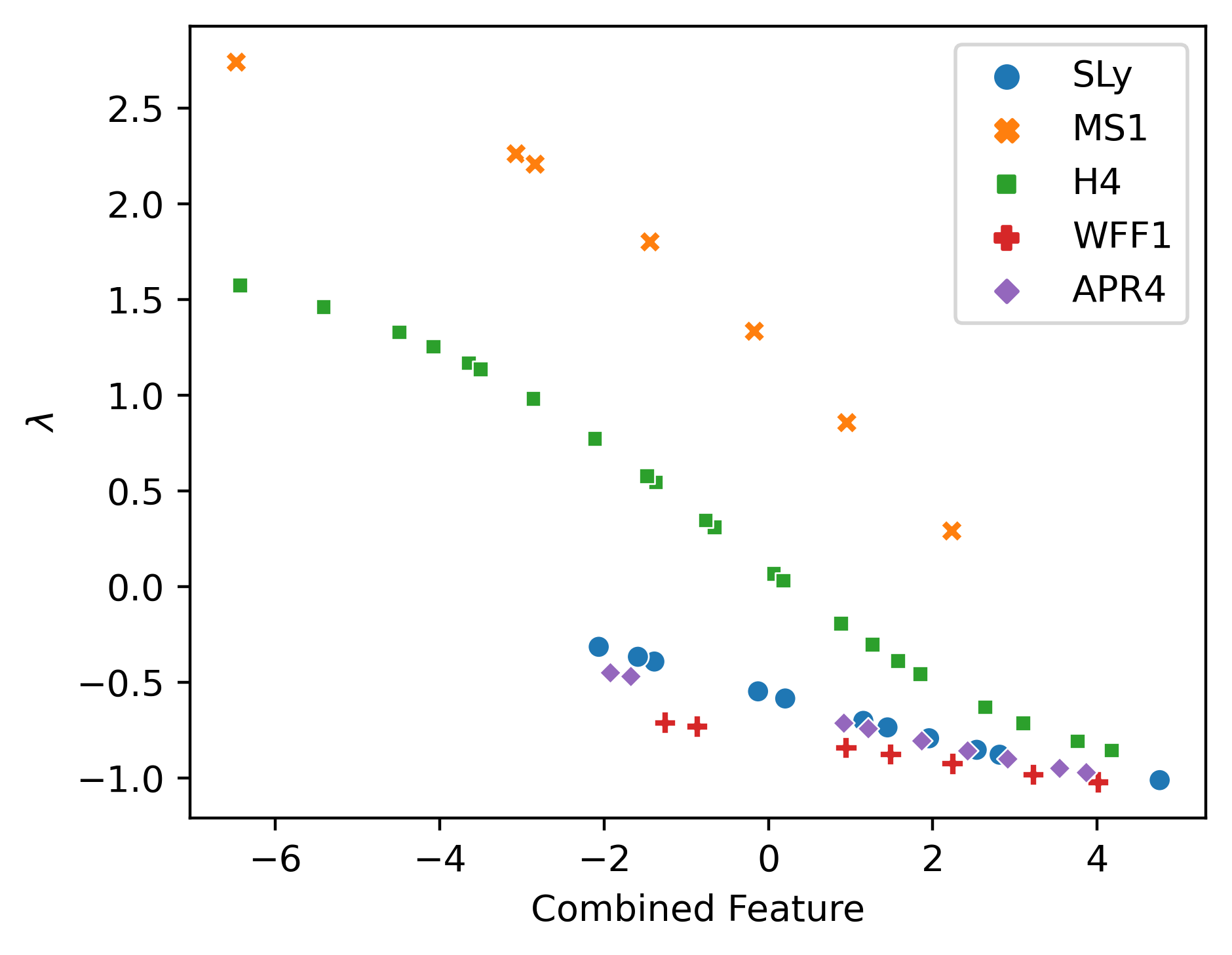}\label{fig:multi2_0}}
\subfloat[][Principal Component 1]{\includegraphics[width=0.5\textwidth]{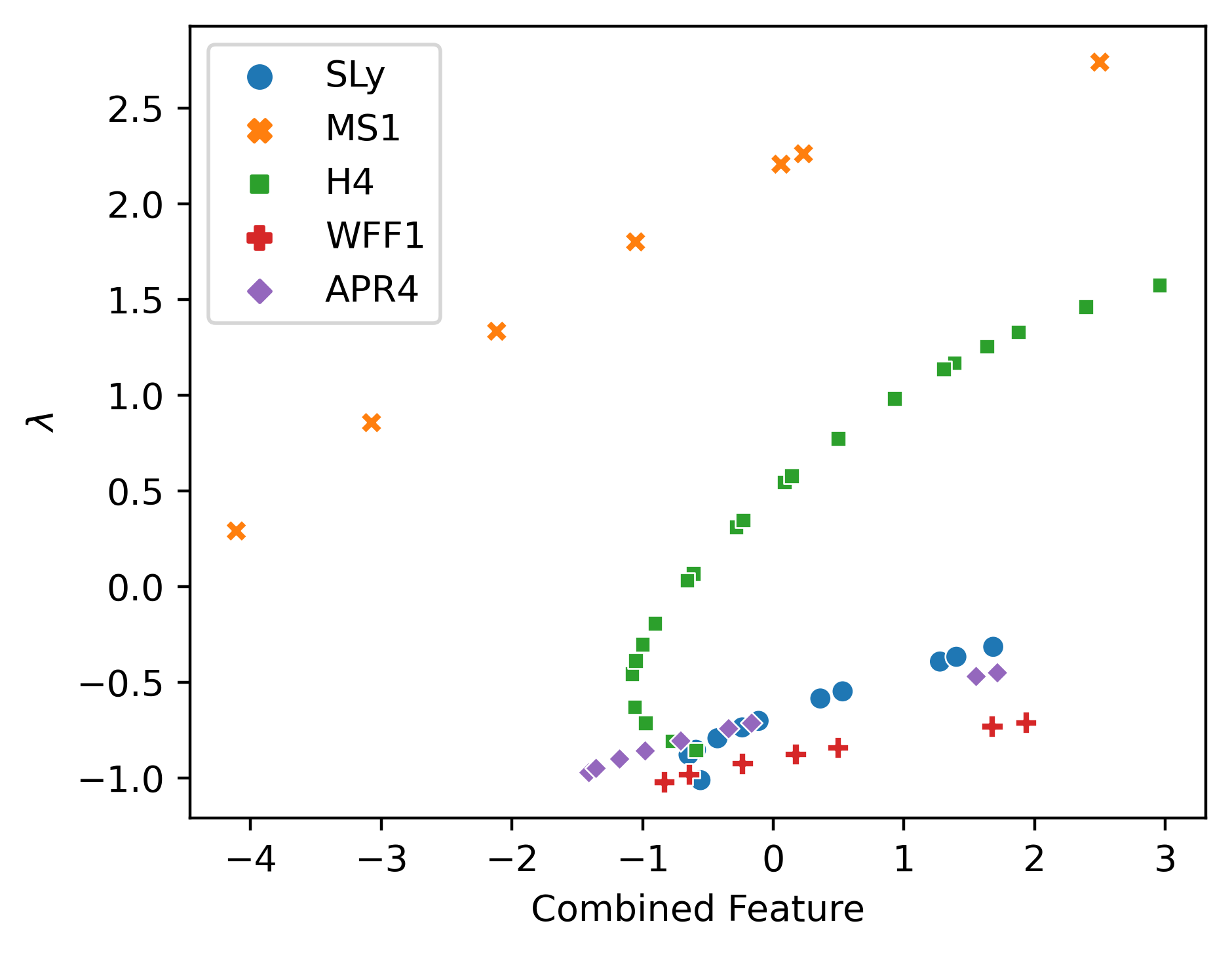}\label{fig:multi2_1}}

\subfloat[][Principal Component 2]{\includegraphics[width=0.5\textwidth]{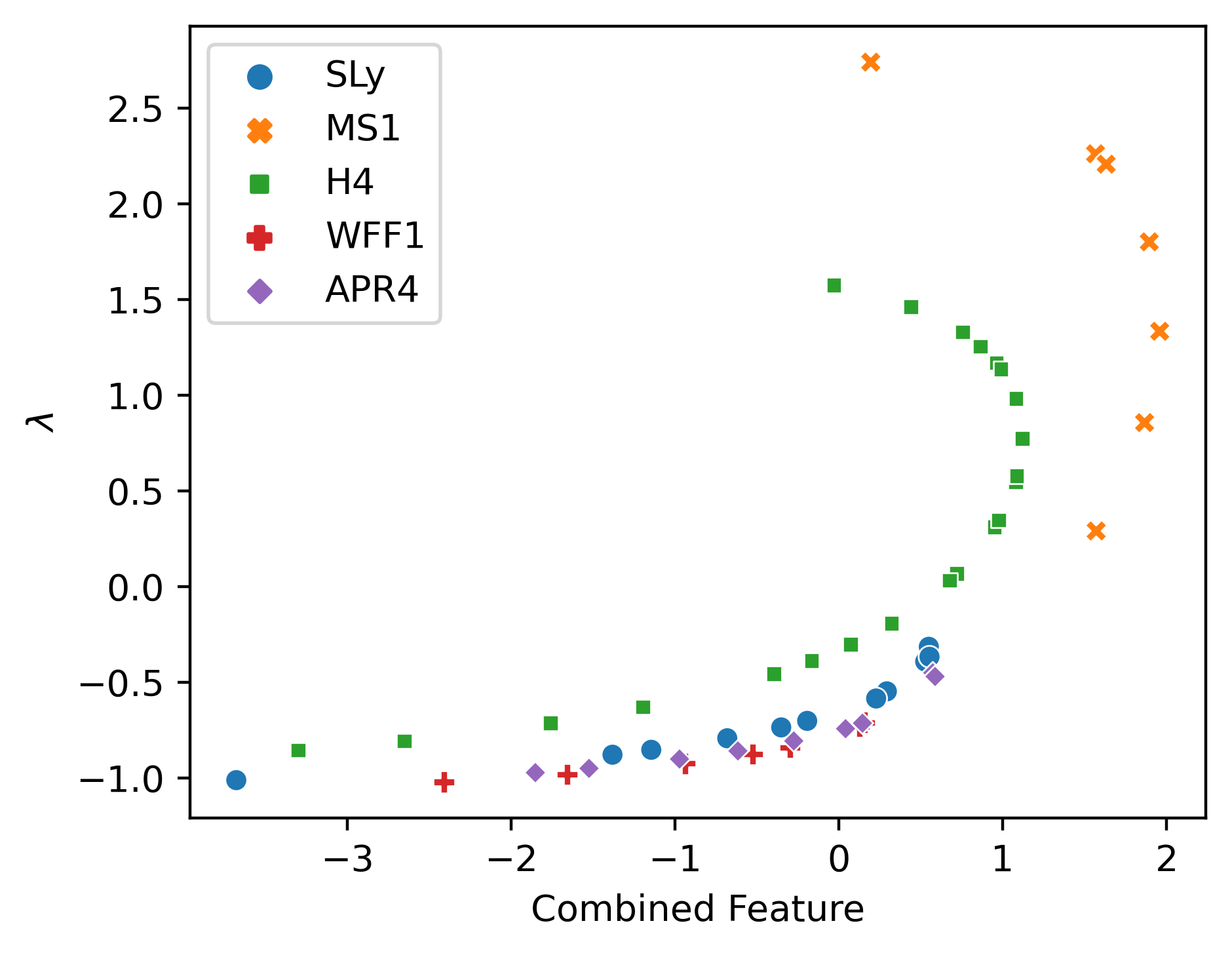}\label{fig:multi2_2}}
\subfloat[][Principal Component 3]{\includegraphics[width=0.5\textwidth]{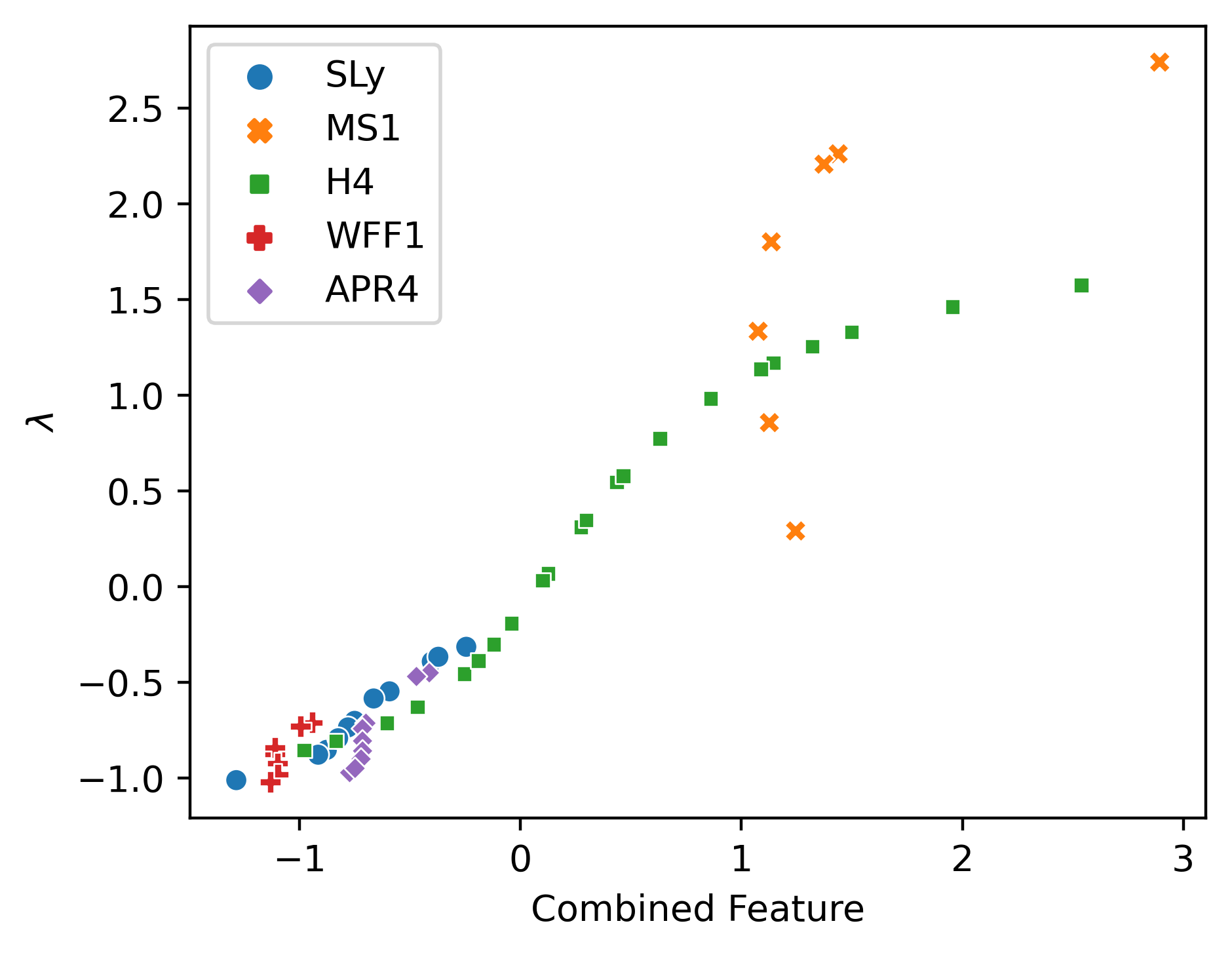}\label{fig:multi2_3}}
\caption{The combined feature derived from each principal component solved for the target feature $\lambda$ after performing PCA on the feature set $\mathbf{F}=\{M, \rho_c, \bar I, \lambda\}$. The corresponding loadings are given in Table~\ref{tab:multi_example2}.}
\label{fig:multi_example2}
\end{figure*}

We now attempt to construct a \AU relation for the \emph{unnormalized} tidal deformablity $\lambda$, using the features $M$, $\rho_c$ and $\bar I$. We again apply the principal component analysis on all 4 features. The resulting principal components are shown in Figure~\ref{fig:multi_example2}. The loadings of each feature corresponding to each principle component are given in Table~\ref{tab:multi_example2}.

As we can clearly see here, none of the combined features derived from the principal components are well correlated with $\lambda$. This is also reflected in the loadings: there is no principal component for which the feature $\lambda$ shows a significantly higher contribution than the other features. 

However, through bivariate analysis, we were previously able to find the well known I-Love~\cite{Yagi365} relation between the \emph{normalized} tidal deformability $\bar \lambda$ and $\bar I$ (cf. Fig~\ref{fig:bivariate_0}). This shows that typically employed normalizations can therefore also not necessarily be overcome by simply employing the PCA approach.

\begin{table}[t]
\def\arraystretch{1.5}%
\centering
\begin{tabular}{c | r | r | r | r }
Component & $\rho_c$ & $M$& $\bar I$ & $\lambda$ \\\hline
0 & \num{-0.51191746} & \num{-0.44792063}& \num{0.53917124} &  \num{0.49659036}   \\
 1 & \num{-0.29858968} & \num{0.77774975} & \num{-0.13208151} & \num{0.53712568} \\
 2 & \num{0.78973118} & \num{0.09031859} & \num{0.44106068}& \num{0.41669256} \\
 3 & \num{-0.15845937} & \num{0.43164219} & \num{0.70520517} & \num{-0.53968631}
\end{tabular}
\caption{Loadings of features in each principal component shown in Figure~\ref{fig:multi_example2}}
\label{tab:multi_example2}
\end{table}

\clearpage
\end{document}